\documentclass[journal]{IEEEtran}
\usepackage{xcolor,soul,framed} 
\usepackage{epstopdf}
\usepackage{lmodern} 
\usepackage{gauss}
\usepackage{amsmath}
\usepackage{lipsum}
\usepackage{graphicx}
\usepackage{subfigure}
\usepackage{setspace}
\usepackage{amssymb}
\usepackage{amsmath}
\usepackage{multirow}
\usepackage{xcolor}
\usepackage{ragged2e}
\usepackage{mathrsfs}
\usepackage{bbold}
\usepackage{float}
\usepackage{stfloats}
\usepackage[all]{xy}

\usepackage[utf8]{inputenc}

\usepackage{tikz}

\usepackage{pgfplots}
\pgfplotsset{compat=newest}
\usetikzlibrary{plotmarks}
\usetikzlibrary{arrows.meta}
\usepgfplotslibrary{patchplots}
\usepackage{grffile}
\usepackage{amsmath}
\usepackage[T1]{fontenc}    
\usepackage{hyperref}       
\usepackage{url}            
\usepackage{booktabs}       
\usepackage{amsfonts}       
\usepackage{nicefrac}       
\usepackage{microtype}      
\usepackage{cite}

\begin{document}

\title{mmWave Spatial-Temporal Single Harmonic Switching Transmitter Arrays for High back-off Beamforming Efficiency}

\author{Zhehao~Yu\textsuperscript{1},
        Xuyang~Lu*,~\IEEEmembership{Member,~IEEE},
        Changzhan~Gu,~\IEEEmembership{Member,~IEEE},
        Suresh~Venkatesh,~\IEEEmembership{Member,~IEEE}, and
        Junfa~Mao,~\IEEEmembership{Fellow,~IEEE}
\thanks{
(Corresponding author: Xuyang~Lu.)
Z. Yu, and X. Lu are with the UM-SJTU Joint Institute - Jiao Tong University, Shanghai 200240, China  (e-mail: Xuyang.Lu@sjtu.edu.cn)

C. Gu, and J. Mao are with Shanghai Jiao Tong University, Shanghai 200240, China

S. Venkatesh is with Princeton University, Princeton, NJ 08544, USA.
}
}

\maketitle

\begin{abstract}
    \textbf{This paper presents a spatial-temporal single harmonic switching (STHS) transmitter array architecture with enhanced efficiency in the power back-off (PBO) region. STHS is an electromagnetic and circuit co-designed and jointly optimized transmitter array that realizes beamforming and back-off power generation at the same time.
    The temporal dimension is originally added in STHS to achieve back-off efficiency enhancement, which can be combined with conventional power back-off enhancement methods such as Doherty amplifiers and envelope tracking. The design is validated through a simulation of a two-stage power amplifier in 65-nm CMOS at 77 GHz, which achieves a peak drain efficiency (DE) of 24.2\%, a 22\% DE at 3-dB PBO, 16\% DE at 6-dB PBO and 10.2\% at 9-dB PBO. The efficiency exhibits a 57\% improvement at 3-dB PBO, 100\% improvement at 6-dB PBO, and 190\% improvement at 9-dB PBO compared with class A/B amplifier.
}
\end{abstract}


\section{Introduction}
Modern wireless communication system takes advantage of high-order modulation schemes to achieve high spectral efficiency by employing dense constellations. One major drawback of transmitting complex modulated signals is the high Peak-to-Average Power Ratio (PAPRs), which forces the transistors to operate in the linear region with a large input back-off. 
As an example, 4G communication uses signals with 7-8 dB power back-off (PBO) while in 5G and WLAN IEEE 802.11ax, the signal PBO can be larger than 9.5 dB   \cite{afaqui2016ieee,cao2020pseudo}. Power amplifiers (PAs) suffer from efficiency degradation when working in the PBO region. 
This raises challenges to wireless system engineers to design highly efficient systems for mobile usage \cite{kazimierczuk2008rf,cripps2006rf}.

\begin{figure*}[h]
    \centering
    \includegraphics[width=0.8\linewidth]{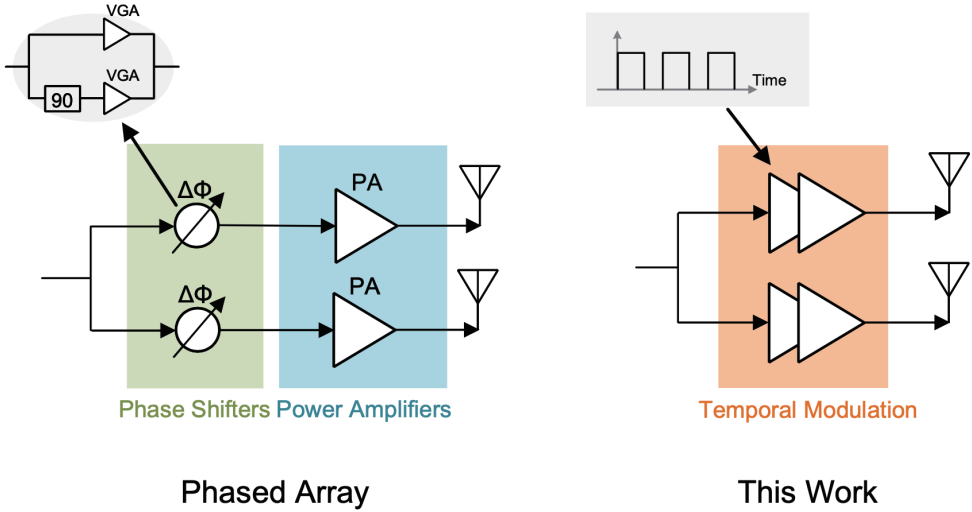}
     \caption{A conventional phased array employs a decoupled approach where the beamforming and power back-off are realized by phase shifters and a power-generating circuit, respectively. In this work, an electromagnetic and circuit co-designed transmitter array is proposed in which a time-modulated array factor and the power generating circuit collaborate to facilitate beamforming and high-efficiency back-off simultaneously. }
     \label{fig:phased_STHS}
\end{figure*}

On the other hand, to get a larger data bandwidth, the carrier frequency will soon enter the mm-Wave band (30-300 GHz).  Limited by the Friis transmission equation, the signal will attenuate with the increased signal carrier frequency. In the mm-Wave band, PAs will show even lower efficiency compared to their low-gigahertz counterparts due to inferior transistor performance at higher frequencies \cite{chowdhury2009design}. For example,
the efficiency of a typical class-A/B power amplifier will degrade from about 60\% at low-gigahertz to lower than 30\% at mm-Wave band. Consequently, to increase the power of the transmitted signal, beamforming techniques are proposed, that is,
 to concentrate the signal transmitted to a certain direction so that the spatial efficiency can be significantly enhanced.
Conventionally, a decoupled approach is applied in designing a wireless system where the beamforming and power back-off is realized by a phased array with phase shifters and a power-generating circuit, respectively.
As shown in Fig. \ref{fig:phased_STHS}, phase shifters with two quadrature paths are implemented in a typical phased array to realize the desired phase shift. On the other hand, a power amplifier is used for controlling the power generation while an efficiency enhancement structure can be used to mitigate PA's efficiency degradation in the back-off region. However, the system efficiency is unsatisfactory because phase shifters introduce extra power loss while power amplifiers' efficiency degrades quickly in the PBO region.
In this work, an electromagnetic and circuit co-designed transmitter array that includes time as an additional degree of freedom is proposed for back-off efficiency enhancement. To introduce the EM-circuit co-designed approach used in this work, antenna array beamforming efficiency and power amplifier back-off efficiency in the conventional approach are first discussed in the following subsections.

\subsection{Beamforming Efficiency in an Antenna Array}
Although phased array offers an additional array gain, it
requires complex weight multiplications in  different signal paths and, consequently, the beamforming efficiency degrades due to poor matching while the system complexity of a phased array is also unappreciated for small-size mobile usage.
To realize beamforming without phase shifters, the concept of time-modulated array (TMA) exploiting time-domain signal processing techniques is proposed \cite{shanks1959four, poli2011harmonic}. 
One challenge faced by TMA design is the presence of  undesired harmonics induced by the periodic time modulation, which results in a harmonic efficiency reduction \cite{bregains2008signal}.
To suppress the sideband level (SBL) while synthesizing desired patterns, different optimization algorithms are proposed, such as  
genetic algorithms \cite{yang2005design}, simulating annealing \cite{fondevila2004optimizing},  particle swarm optimization \cite{poli2010handling}, and differential evolution \cite{zhu2012design}. 
An abundance of works can be found in the literature where different applications of the time-modulated array are proposed.
Li et al. studied TMA from the view of array factors to realize adaptive beamforming \cite{li2010hybrid} and direction finding \cite{li2010direction}.
Masotti et al. took  advantage of TMA's real-time beamforming features in the application of wireless power transfer \cite{masotti2016time}.
Also, a system-level design is further proposed by Yao et al. that TMA could be applied together with I/Q modulation to generate a scanning beam at a single positive sideband while suppressing other harmonics \cite{yao2015single}.

\subsection{Back-off Efficiency for Power Amplifiers}
After introducing the beamforming efficiency in antenna arrays, the back-off efficiency for  power amplifiers is discussed in this subsection.
 Multiple techniques have been proposed to enhance the efficiency of PAs while transmitting  high PAPR signals including envelope-tracking architecture, Doherty, and outphasing. The essence of those methods is to preserve the operating conditions of power amplifiers in their high-efficiency conditions including bias voltages and impedances and to reduce DC power consumption when operating in the back-off region. The relative pros and cons among those circuit-level back-off efficiency enhancement architectures are summarized in Table \ref{table:PAtable}.

Envelope-tracking PA controls the supply voltage of the PA according to the envelope of the transmitted signal to optimize the efficiency of PAs \cite{kang2013envelope, mahmoudidaryan2019wideband}. In this case, systems bandwidth is traded for higher efficiency.
The Doherty PA is another popular structure that takes  advantage of the load-pull realized by the main PA and the auxiliary PA \cite{kim2006doherty, camarchia2015doherty}. When power is delivered at back-off, the auxiliary PA will be turned off, and for increased signal power, the load-pull techniques retain the impedance seen by the main PA at its optimal value. However, the necessity of complex output matching inevitably introduces a loss penalty of about 1.7 dB in a 2 GHz range and this matching loss will increase as the carrier frequency increases.
A typical Doherty PA implemented operating at 77 GHz in a 40-nm CMOS process exploits a drain efficiency enhancement of 50\% at 3-dB PBO, 100\% at 6-dB PBO compared with a typical class-A/B \cite{kaymaksut2015transformer}.
Outphasing PA uses two PAs driven by signals with different phases so that the system could deliver the desired output power \cite{raab1985efficiency,godoy20122,tai2012transformer}. Whereas, it has a limited output power dynamic range.

Subharmonic switching is an alternative technique that utilizes the lower magnitude of high-frequency harmonics of the modulation pulses of a class-D power amplifier \cite{zhang2019subharmonic}. The power back-off is realized by lowering the switching frequency of a class-D digital PA to utilize subharmonics of the output signals. Yet, since the relative magnitude of subharmonics is fixed, PA branches need to be used to achieve different power back-off regions, which results in higher system complexity. Also, digital PAs used in subharmonic switching will fail in high-frequency operations due to the low quality factor of the output filtering network.

\begin{table*}[h]
\centering
\caption{Comparison between circuit-level back-off efficiency enhancement architectures }
\begin{tabular}{c|c|c}
\hline
\hline
\textbf{Architecture}  & \textbf{Pros}  & \textbf{Cons}                                               \\
\hline
\multirow{2}{*}{\textbf{Envelope-tracking} \cite{kang2013envelope}} 
& Optimal efficiency at   & High bandwidth supply needed.    \\
   & the deep PBO regions.   & Accurate envelope signals required.   \\
 \hline
\multirow{2}{*}{\textbf{Doherty} \cite{kim2006doherty}}    & \multirow{2}{*}{Relatively high efficiency.} & \multirow{2}{*}{Complex output matching networks.} \\
   &         &                          \\
\hline
\multirow{2}{*}{\textbf{Outphasing} \cite{tai2012transformer}}            & \multirow{2}{*}{Low system complexity.}       & Low bandwidth.                                     \\
    & & Limited dynamic power range.   \\
    \hline
\multirow{2}{*}{\textbf{Subharmonic switching} \cite{zhang2019subharmonic}} & \multirow{2}{*}{High deep PBO efficiency.}    & PA branches needed.                                \\
 &   & Low-frequency application only.   \\
\hline
\hline
\end{tabular}
\label{table:PAtable}
\end{table*}

\begin{figure*}[h]
    \centering
\subfigure[]{
\includegraphics[width=0.48\linewidth]{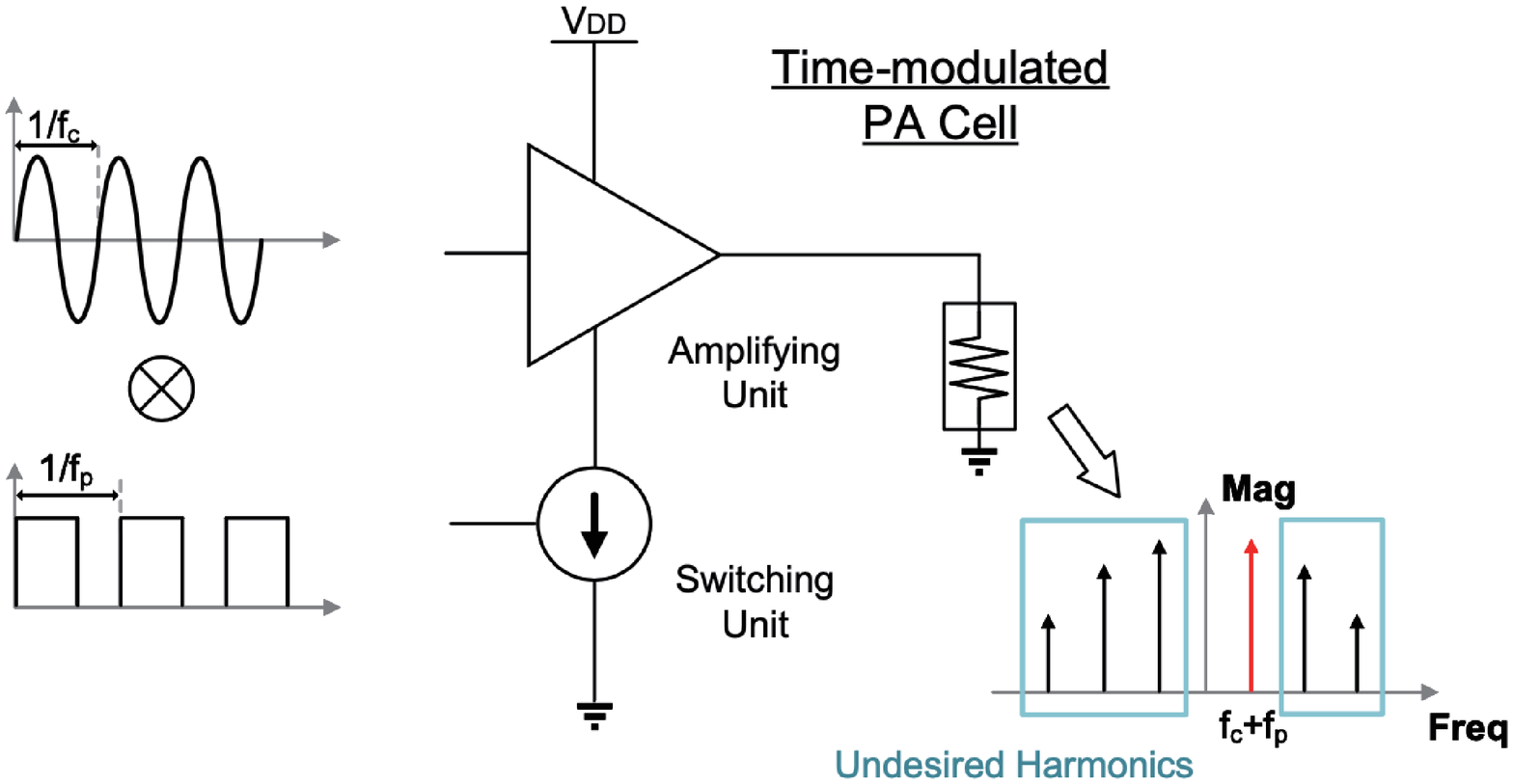}
}
\subfigure[]{
\includegraphics[width=0.48\linewidth]{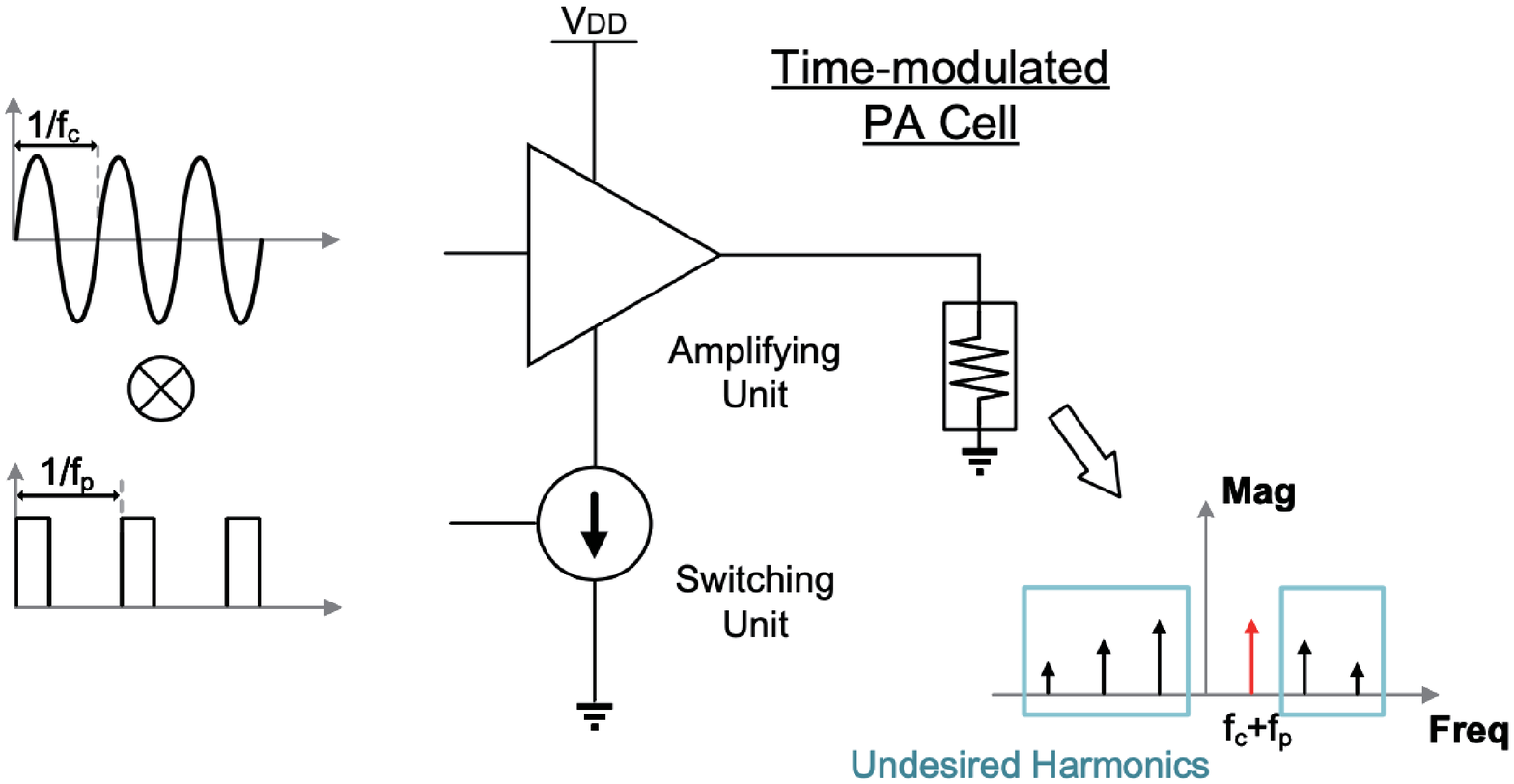}
}
\caption{A time-modulated PA cell in (a) the peakmode, (b) the pbomode. The first harmonic at  $f_c+f_p$
is utilized to deliver signals while the power back-off can be achieved by decreasing the duty cycle of the modulating pulses. However, the harmonic efficiency degenerates significantly because of the existence of other harmonics induced by the modulating pulses.}
\label{fig:TimemodulatedPA}
\end{figure*}


\subsection{EM-circuit Co-design}
As illustrated in Fig. \ref{fig:phased_STHS}, since the decoupled design is used in a conventional phased array, power amplifiers and phase shifters are optimized separately before they are combined. 
In this work, an electromagnetic and circuit co-designed transmitter array is proposed that realizes  spatial beamforming and power back-off efficiency enhancement by exploiting the synergy between circuits and the array.
Since this system makes use of the first harmonic signal of the temporal switching pulse, we refer to it as spatial-temporal single harmonic switching (STHS) transmitter array in the following discussion.

 It is first noted that the duty cycle of time-domain pulse modulation is viable to control the output signal power. In the ideal case, the PA switches ON and OFF according to the modulating pulses, which is similar to the operation of a class-D power amplifier, leading to an improved efficiency. This shows the potential to achieve beamforming and enhance  power back-off efficiency at the same time by combining an  antenna array with controllable pulse modulation. However, the energy will be wasted on undesired harmonics induced by the  pulse modulation. To deal with this power loss, a carefully designed time-modulated phased array switching sequence is used to improve array beamforming efficiency for both the peakmode and the power back-off mode.
By the adoption of the time domain pulse modulation signal in an antenna array, the power on undesired harmonics will be suppressed spatially resulting in a system-level back-off efficiency enhancement.
It is also important to notice that, since the STHS optimizes the array factor in the spatial and the temporal domain, additional system efficiency could be further achieved by combining STHS with existing circuit level optimization methods such as the Doherty PA.

The remainder of this paper is organized as follows. 
The co-design considerations for the proposed STHS array are presented in Section II where the system theoretical model is derived. 
Section III presents the simulated characteristics for the STHS array.
Finally, the conclusion is drawn in Section VI.

\section{Co-Design and Optimization}
\subsection{Proposed Spatial-Temporal Single Harmonic Switching (STHS) Array}

\begin{figure*}[h]
    \centering
    \includegraphics[width=0.8\linewidth]{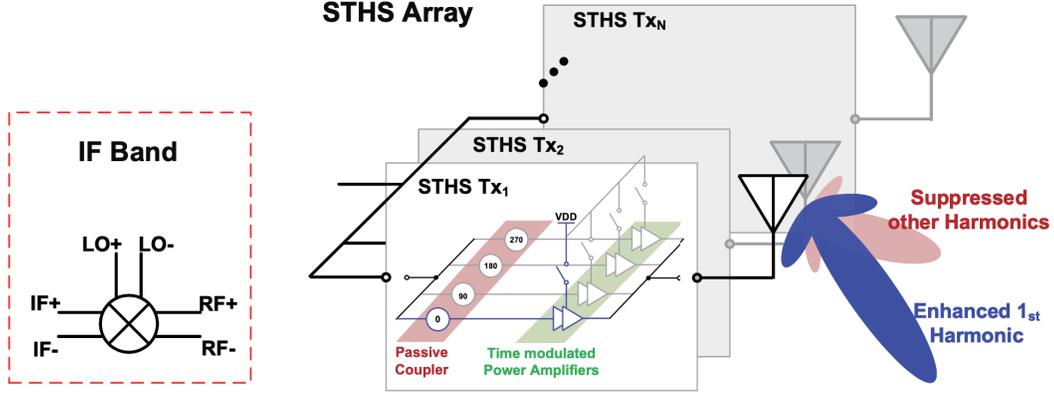}
    \caption{Architecture of an STHS transmitter array.}
    \label{fig:TMAconfig}
\end{figure*}
\begin{figure}[h]
\centering
\includegraphics[width=1\linewidth]{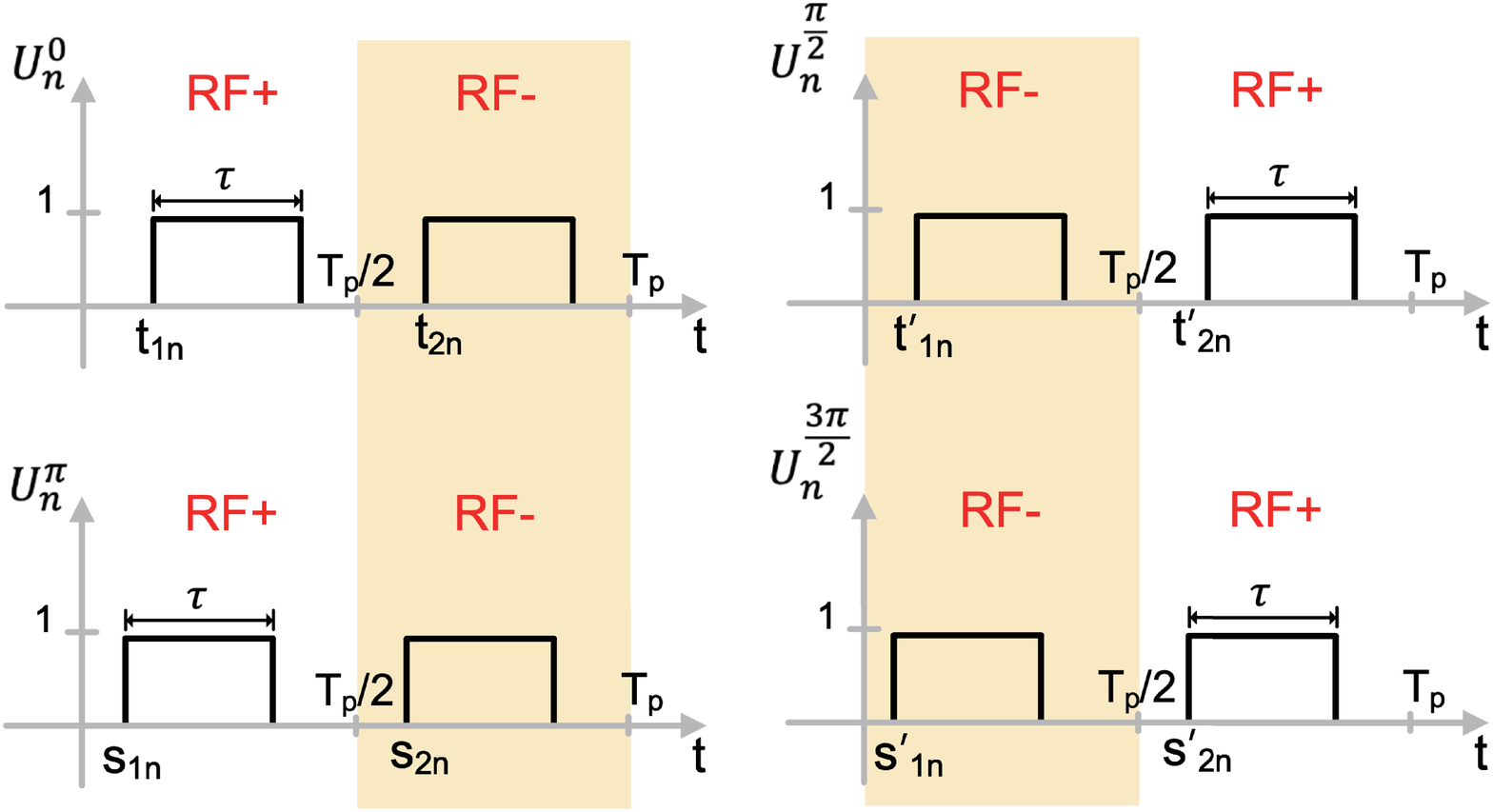}
\caption{Modulation pulses for 4 signal paths with $90^{\circ}$ phase differences
 ($U^{0}_{n}$,
 $U^{-\frac{\pi}{2}}_{n}$, 
$U^{-\pi}_{n}$, $U^{-\frac{3\pi}{2}}_{n}$) at the $n^{th}$ array element.
}
\label{fig:pulse}
\end{figure}

Generally speaking, the PA output impedance is optimized for maximal power efficiency at the peakmode. When the PA goes into the back-off region, the efficiency will degrade since the output power decreases while the DC power remains nearly the same. One way to improve the back-off efficiency is to let the PA either work in the peakmode or be turned off.  This concept is illustrated in Fig. \ref{fig:TimemodulatedPA} where
a power amplifier is connected to a voltage-controlled current source modulated by the switching pulse sequence controlling the power generation. Ideally, the PA will work with maximum efficiency when the pulse is high and will be turned off when the pulse is low. The maximum output power is generated when the time-modulated PA cell works in peakmode while the back-off power is delivered in pbomode by decreasing the duty cycle of the modulating pulse. The time domain multiplication between the transmit signal and modulating pulse will result in a frequency domain convolution, and therefore, the frequency spectrum of the output signal will consist of all the odd harmonics.
Although each harmonic could be utilized to deliver the output signal theoretically, the bandwidth is often restricted in the modern communication systems to avoid the cross-talk between channels. In this work, we utilize the first harmonic to deliver signals since it possesses more power as compared with other harmonics.
To show how the back-off power is delivered by the modulating pulse, the first harmonic at $f_c+f_p$ is considered
where $f_c$ denotes the carrier frequency and  $f_p$ denotes the modulating pulse frequency.
The $f_c$ and $f_p$ should satisfy $f_c \gg f_p$ and $f_p > 2f_{IF}$
where $f_{IF}$ denotes the intermediate frequency to avoid aliasing.
Since the duty cycle determines the duration for the power amplifier to be turned on, it is seen that a decrease in the duty cycle of the modulating pulse signal will reduce the output magnitude of the first harmonic at $f_c+f_p$.

The overall efficiency of the co-designed transmitter array is determined by both the harmonic efficiency $\zeta_{harm}$ of the time-modulated array and the drain efficiency of the circuit $\zeta_{circ}$. Hence, the transmitter array drain efficiency (DE) $\eta$ could be expressed as 
\begin{equation}
\eta =\zeta_{harm} \times \zeta_{circ}.
\label{Eq:systemeff}
\end{equation}
If we denote $P_{(m)}$ 
as the power of the $m^{th}$ harmonic
 where $m\in \mathbb{Z}$,
the array harmonic efficiency $\zeta_{harm}$ being the ratio between the power of the desired first harmonic ($f_c+f_p $), and the total power radiated $P_{tot}$
can be written as
\begin{equation}
    \zeta_{harm} = \frac{P_{(1)}}{P_{(tot)}}.
\end{equation}
In Fig. \ref{fig:TimemodulatedPA}, it is seen that the undesired harmonics ($f_c+3f_p$, $f_c-f_p$, etc.) will lead to an efficiency degradation. To improve the harmonic efficiency, a properly designed time-modulated pulse eliminating undesired harmonics will be implemented and discussed.

In terms of the circuit efficiency $\zeta_{circ}$, two non-idealities need to be dealt with that degrades the back-off efficiency:
\begin{enumerate}
    \item When the PA cell is in the "OFF" state, the existence of sub-threshold leakage power degrades the efficiency.
    \item During each switching period, there exist dynamic power losses resulting from the charging and discharging process of the parasitic capacitance.
\end{enumerate}
To optimize the circuit efficiency, the efficiency model for the power loss is established and 
circuit parameters like the size of transistors are optimized to balance the trade-off between switching performance and efficiency degradation.

\subsection{Harmonic Efficiency Optimization through STHS}
In this subsection, the configuration of a 4-path STHS array shown in Fig. \ref{fig:TMAconfig} is discussed to improve array harmonic efficiency and to realize beamforming. The STHS array utilizes the differential RF signal pair ($RF+$ and $RF-$) generated by a double-balanced mixer to realize a pulse modulation with different polarities.
Each transmitter is composed of four signal paths with $90^{\circ}$ phase differences generated by a quadrature hybrid. Each signal path consists of a time-modulated PA cell to realize independent pulse modulation. The output signal after modulation is recombined and delivered to each antenna element.

As discussed before, the switching waveform contains different harmonics while only the first harmonic will be used in this design for spectral-efficient beamforming. Therefore, higher efficiency can be achieved by carefully designing the time-modulated pulses such that the first positive harmonic frequency is enhanced while the other harmonics are suppressed. 
According to \cite{yao2015single}, multiple modulating pulses can be designed such that a combination of signals modulated by those pulses will lead to suppression of undesired harmonics.
The modulation pulses for each path are illustrated in Fig. \ref{fig:pulse}. If we assume that all the elements are isotropic and equally spaced, the array factor can be written as
\begin{equation}
\begin{split}
AF&(\theta, t)=e^{j 2 \pi f_{0} t} \sum_{n=1}^{N} I_{n} \\
\cdot&\left[U^{0}_{n}(t)+U^{-\frac{\pi}{2}}_{n}(t)+
U^{-\pi}_{n}(t)+U^{-\frac{3\pi}{2}}_{n}(t)\right] \cdot e^{j \beta(n-1) d \sin \theta}
\end{split}
\end{equation}
where $f_{0}$ is the carrier frequency, $T_p$ 
is the time-modulated pulse period assuming that $f_0\gg f_p$ ($f_p = 1/T_p$). $\beta = 2\pi/\lambda$ is the free-space wave number, $\lambda = c/f_0$ is the wavelength, and c is the speed of light in vacuum. 
$I_0$ is the amplitude excitation of the  $n^{th}$ element. $\theta$ is the angle observed from the broadside direction, and $d$ denotes the interelement distance. $U^{0}_{n}(t)$, $U^{-\frac{\pi}{2}}_{n}(t)$, $U^{-\pi}_{n}(t)$, $U^{-\frac{3\pi}{2}}_{n}(t)$ are  the modulating pulses on  signal paths with different phases used to modulate the $n^{th}$
array element.
To study the signal in the frequency domain, the pulse functions are expressed through Fourier series with different frequency components
\begin{equation}
\left\{
\begin{aligned}
U^{0}_{n}(t) &=\sum_{m=-\infty}^{+\infty} a^{0}_{m n} \cdot e^{j 2 \pi m \cdot f_{p} t}, \quad m \in \mathbb{Z} \\
U^{-\frac{\pi}{2}}_{n}(t) &=\sum_{m=-\infty}^{+\infty} a^{-\frac{\pi}{2}}_{m n}\cdot e^{j 2 \pi m \cdot f_{p} t}, \quad m \in \mathbb{Z}\\
U^{-\pi}_{n}(t) &=\sum_{m=-\infty}^{+\infty} a^{-\pi}_{m n}\cdot e^{j 2 \pi m \cdot f_{p} t}, \quad m \in \mathbb{Z}\\
U^{-\frac{3\pi}{2}}_{n}(t) &=\sum_{m=-\infty}^{+\infty} a^{-\frac{3\pi}{2}}_{m n}\cdot e^{j 2 \pi m \cdot f_{p} t}, \quad m \in \mathbb{Z}
\end{aligned}
\right.
\end{equation}
where $\mathbb{Z}$ is the set of all integers.

Then the array factor becomes
\begin{equation}
\begin{aligned}
A&F(\theta, t)= \sum_{m=-\infty}^{+\infty} e^{j 2 \pi\left(f_{0}+m \cdot f_{p}\right) t} \\
\cdot& \sum_{n=1}^{N} I_{n} \cdot\left(a^{0}_{m n}+a^{-\frac{\pi}{2}}_{m n}+a^{-\pi}_{m n}+a^{-\frac{3\pi}{2}}_{m n}\right) \cdot e^{j \beta(n-1) d \sin \theta}.
\end{aligned}
\end{equation}
In the following calculation, given the pulse frequency $f_p$, parameters to be designed include:
the duration of each modulating pulse $\tau$ and the switch-ON time of positive and negative pulses for each path, i.e., $t_{1n}\text{, } t_{2n}\text{ for } U^0_n$, $t_{1n}^{\prime}\text{, } t_{2n}^{\prime}\text{ for } U^{-\frac{\pi}{2}}_n$, 
$s_{1n}\text{, } s_{2n}\text{ for } U^{-\pi}_n$,
and $s_{1n}^{\prime}\text{, } s_{2n}^{\prime}\text{ for } U^{-\frac{3\pi}{2}}_n$.

To completely suppress the carrier component, the modulation pulses should have a mean of 0. This means that the duration of $RF+$ and $RF-$ should be equal so that the Fourier coefficients for the carrier component will be zero, i.e.,
\begin{equation}
a_{0n}^0=a_{0n}^{-\frac{pi}{2}}=a_{0n}^{\pi}=a_{0n}^{\frac{3\pi}{2}} =0,
\end{equation}
For $m\neq 0$,
the Fourier coefficient for $U^0_n$  can be expressed as

\begin{equation}
\begin{aligned}
&a_{m n}^{0} =\frac{1}{T_{p}} \int_{0}^{T_{p}} U_{n}^{0}(t) e^{-j 2 \pi m \cdot f_{p}} d t \\
&=\frac{2}{m \pi} \sin \left(m \pi \frac{\tau}{T_{p}}\right) \sin \left(m \pi \frac{t_{2 n}-t_{1 n}}{T_{p}}\right) e^{-j m \pi \frac{t_{2 n}+t_{1 n}+\tau}{T_{p}}} e^{j \frac{\pi}{2}} .
\end{aligned}
\end{equation}
A similar calculation can be applied to the rest three paths ($a^{-\frac{\pi}{2}}_{m n}\text{, } a^{-\pi}_{m n}\text{, and } a^{-\frac{3\pi}{2}}_{m n}$). Then the array factor is written as
\begin{equation}
A F_{m}(\theta, t)=e^{j 2 \pi\left(f_{0}+m \cdot f_{p}\right) t} \cdot \sum^{N} I_{n} A_{m n} e^{j \beta(n-1) d \sin \theta}
\end{equation}
where
\begin{equation}
A_{m n}=a^{0}_{m n}+a^{-\frac{\pi}{2}}_{m n}+a^{-\pi}_{m n}+a^{-\frac{3\pi}{2}}_{m n}.
\end{equation}

\begin{figure*}[h]
    \centering
\subfigure[]{
\includegraphics[width=0.48\linewidth]{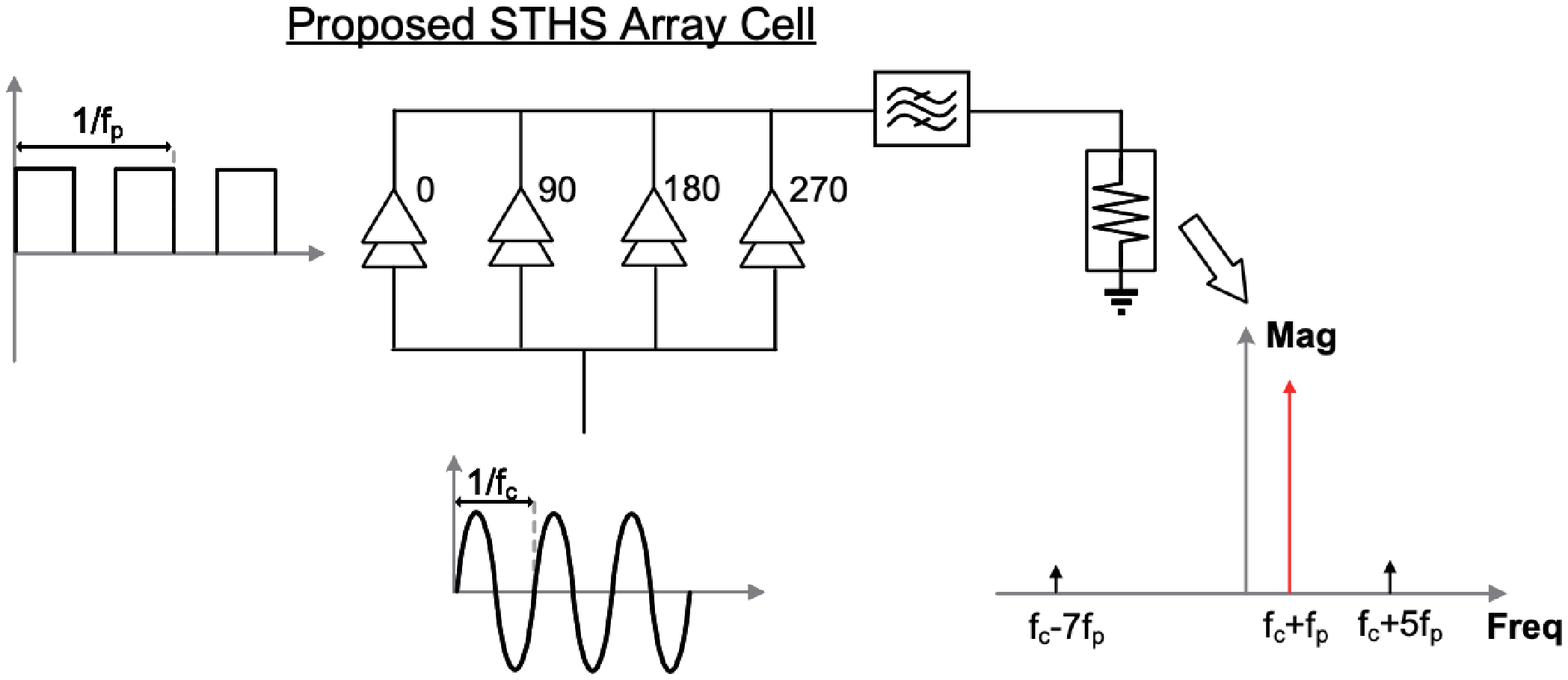}
}
\subfigure[]{
\includegraphics[width=0.48\linewidth]{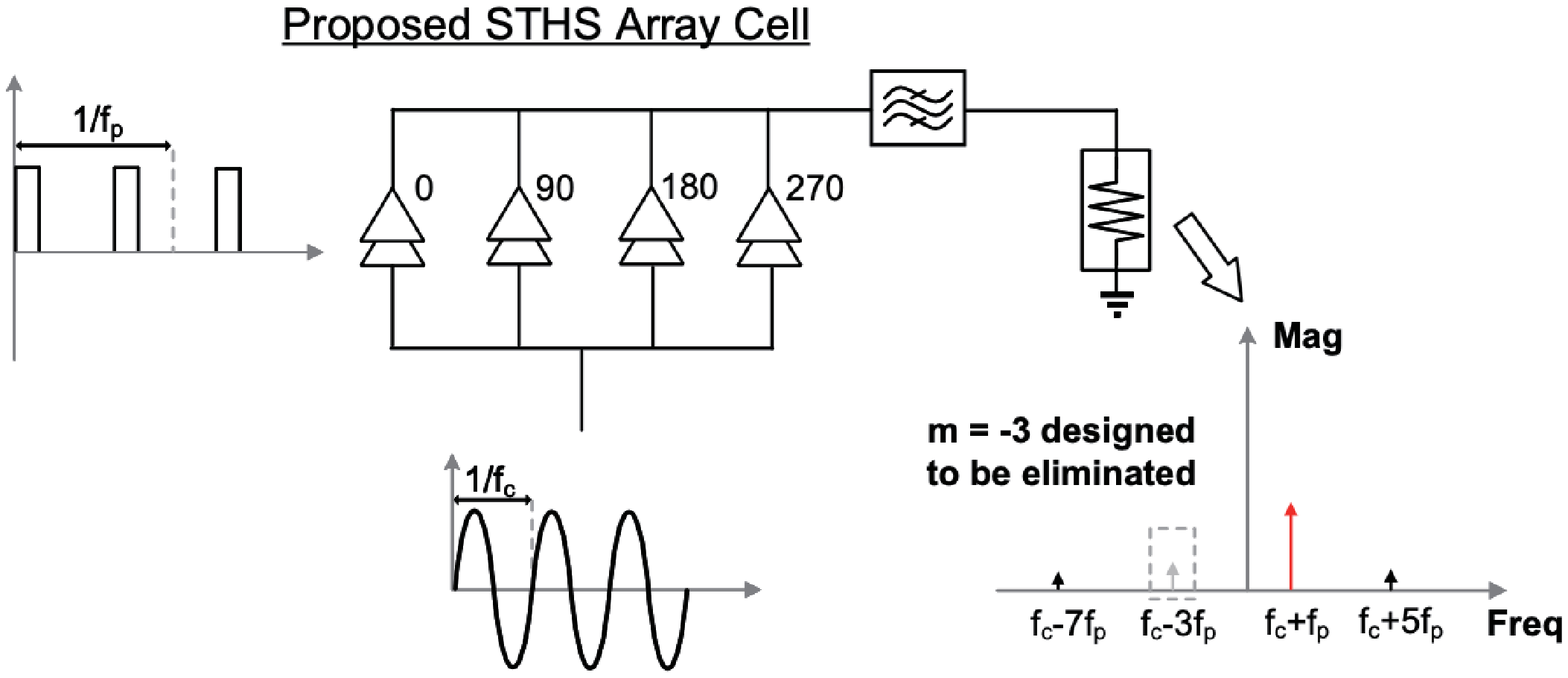}
}
\caption{An STHS transmitter array cell in (a) the peakmode, (b) the pbomode. The first harmonic at $f_c+f_p$
is utilized to deliver signals while the power back-off signal is delivered by decreasing the duty cycle of modulating pulses. Those pulses after modulation along four paths are combined to suppress other undesired harmonics. In the pbomode, the $m = -3$ harmonic is suppressed according to the designed modulating pulses.}
\label{fig:STHSPA}
\end{figure*}

As for the array harmonic efficiency
\begin{equation}
\zeta_{harm}= \frac{P_{(1)}}{P_{(tot)}}=\frac{P_{(1)}}{\sum_{m=-\infty}^{m=\infty}P_{(m)}}.
\label{Eq:tmaEff}
\end{equation}
According to Parseval's theorem, the total power from the $m^{th}$ harmonic $P_{(m)}$ is given by
\begin{equation}
P_{(m)}=\sum_{n=1}^{N} \sum_{s=1}^{N}\left\{I_{n} I_{s} \operatorname{sinc}[\beta d(n-s)] \cdot A_{m n} A_{m s}^{*}\right\},
\label{Eq:harmonicsPower}
\end{equation}
which indicates that $A_{mn}$ can be designed to optimize
harmonic efficiency by enhancing the power delivered through the first positive harmonic. 
In the following calculation, harmonics with 
$m \in \{2k\text{, } 3k\text{, }4k-1, \quad  k \in \mathbb{Z} \}$ 
are designed to be eliminated.

To eliminate even harmonics, the coefficients should satisfy
\begin{equation}
\left\{
\begin{aligned}
\frac{t_{2 n}-t_{1 n}}{T_{p}}&=\frac{t_{2 n}^{\prime}-t_{1 n}^{\prime}}{T_{p}}=\frac{1}{2},\\
\frac{s_{2 n}-s_{1 n}}{T_{p}}&=\frac{s_{2 n}^{\prime}-s_{1 n}^{\prime}}{T_{p}}=\frac{1}{2}.
\end{aligned}
\right.
\end{equation}
Then $A_{mn}$
can be reformed into:
\begin{equation}
\begin{aligned}
&A_{m n}=\frac{2}{m \pi} \sin \left(m \pi \frac{\tau}{T p}\right) \sin \left(\frac{m \pi}{2}\right) \operatorname{sgn}(m)\\
&\cdot\left[e^{-j m \pi \frac{2 t_{1 n}+\tau}{T_{p}}} 
e^{j \pi \frac{1-m}{2}}+e^{-j m \pi \frac{2t_{1n}^{\prime}+\tau}{T_{p}}} e^{j \pi \frac{-m}{2}}+\right. \\
&\left.e^{-j m \pi \frac{2 s_{1 n}+\tau}{T_{p}}} e^{j\pi\frac{-1-m}{2}}+
e^{-j m \pi \frac{2 s_{1n}^{\prime}+\tau}{T_{p}}} 
e^{j \pi \frac{-2-m}{2}}\right]
\end{aligned}
\label{Eq:Amn}
\end{equation}
where $\operatorname{sgn}(m)$ denotes the sign of parameter $m$.
It is seen that all even harmonics are eliminated, i.e.,
\begin{equation}
\left|A_{m n}\right|=0, \quad\{m=2 k, \quad k \in \mathbb{Z}\}
\end{equation}
Similarly, the pulse duration can be set as
\begin{equation}
\frac{\tau_{0}}{T_{p}}=\frac{1}{3}
\label{Eq:Tp}
\end{equation}
to suppress all $m^{th}$ harmonics with $m=3k\text{, } k \in \mathbb{Z}$, i.e.,
\begin{equation}
\left|A_{m n}\right|=0, \quad\{m=3k, \quad k \in \mathbb{Z}\}.
\end{equation}
To eliminate harmonics with $m = 4k - 1$  with $ k \in \mathbb{Z}$, $t_{1 n}^{\prime}$ and $s_{1 n}^{\prime}$ should satisfy
\begin{equation}
\left\{
\begin{aligned}
\frac{t_{1 n}^{\prime}}{T_{p}}&=\frac{t_{1 n}}{T_{p}}-\frac{1}{4},\\
\frac{s_{1 n}^{\prime}}{T_{p}}&=\frac{s_{1 n}}{T_{p}}-\frac{1}{4}.
\end{aligned}
\right.
\label{Eq:4k-1}
\end{equation}
Then   (\ref{Eq:Amn}) could be rewritten as
\begin{equation}
\begin{aligned}
A_{m n}&=\frac{4}{m \pi} \sin \left(m \pi \frac{\tau}{T p}\right) \sin \left(\frac{m \pi}{2}\right) \operatorname{sgn}(m) \\
\cdot &\left[e^{-j m \pi \frac{2 t_{1 n}+\tau}{T_{p}}} e^{j \pi \frac{1-m}{2}}+
e^{-j m \pi \frac{2 s_{1 n}+\tau}{T_{p}}} e^{j \pi \frac{-1-m}{2}}\right].
\end{aligned}
\label{Eq:AmnSimplified}
\end{equation}

\begin{figure*}[h]
\begin{minipage}[h]{0.48\linewidth}
     \input{Figs/PBO0_theta20.tex}
     \caption{Ideal radiation patterns  
     with scanning angle
    $\theta=20^{\circ}$ in the peakmode.}
     \label{fig:ideal0}
  \end{minipage}%
  \hspace{5mm}
  \begin{minipage}[h]{0.48\linewidth}
      \input{Figs/PBO6_theta20.tex}
     \caption{Ideal radiations pattern  
     with scanning angle
    $\theta=20^{\circ}$ with a duty cycle ratio of $10log(\alpha)=  -6$. The $m=-3$ harmonic is suppressed.
    Ideally, this corresponds to a 6-dB PBO.}
     \label{fig:ideal6}
  \end{minipage}%
   \end{figure*}

\begin{figure*}[h]
    \centering
\subfigure[]{
\includegraphics[width=0.31\linewidth]{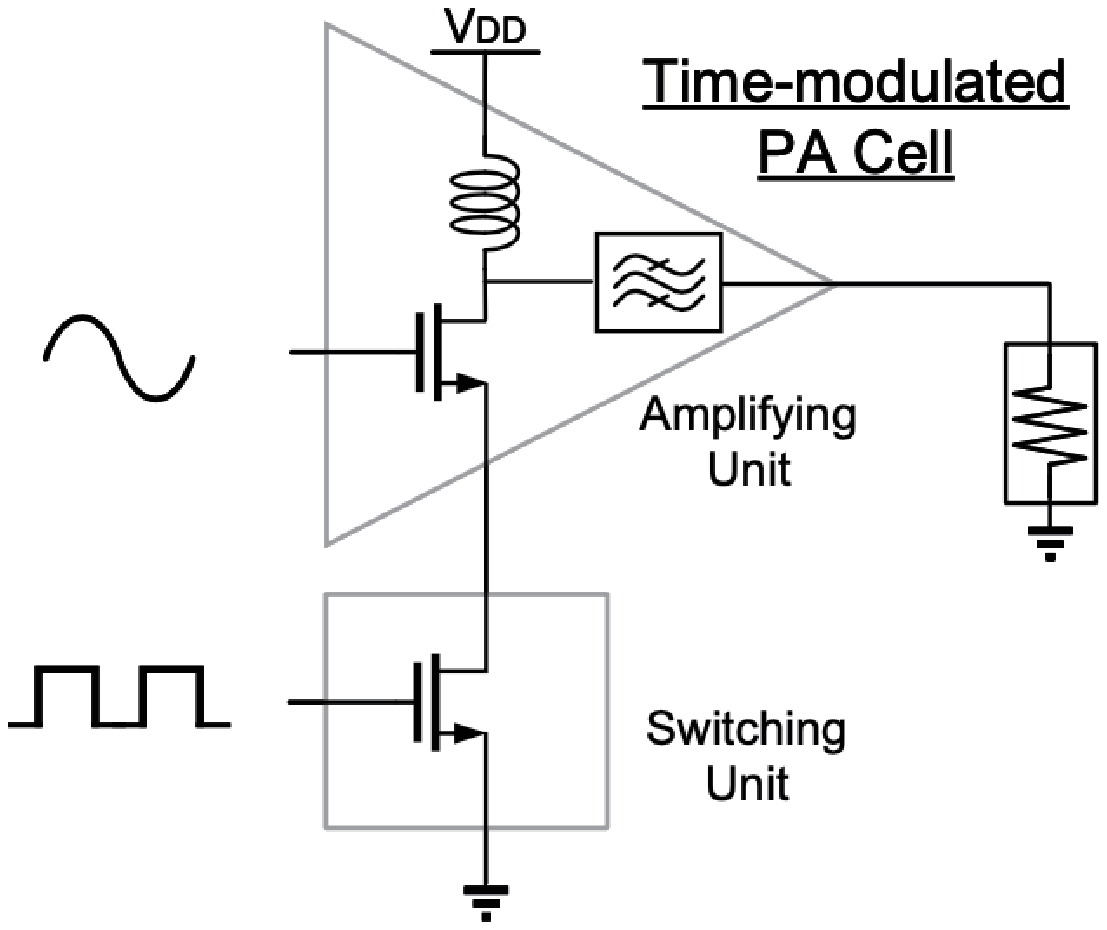}
}
\subfigure[]{
\includegraphics[width=0.31\linewidth]{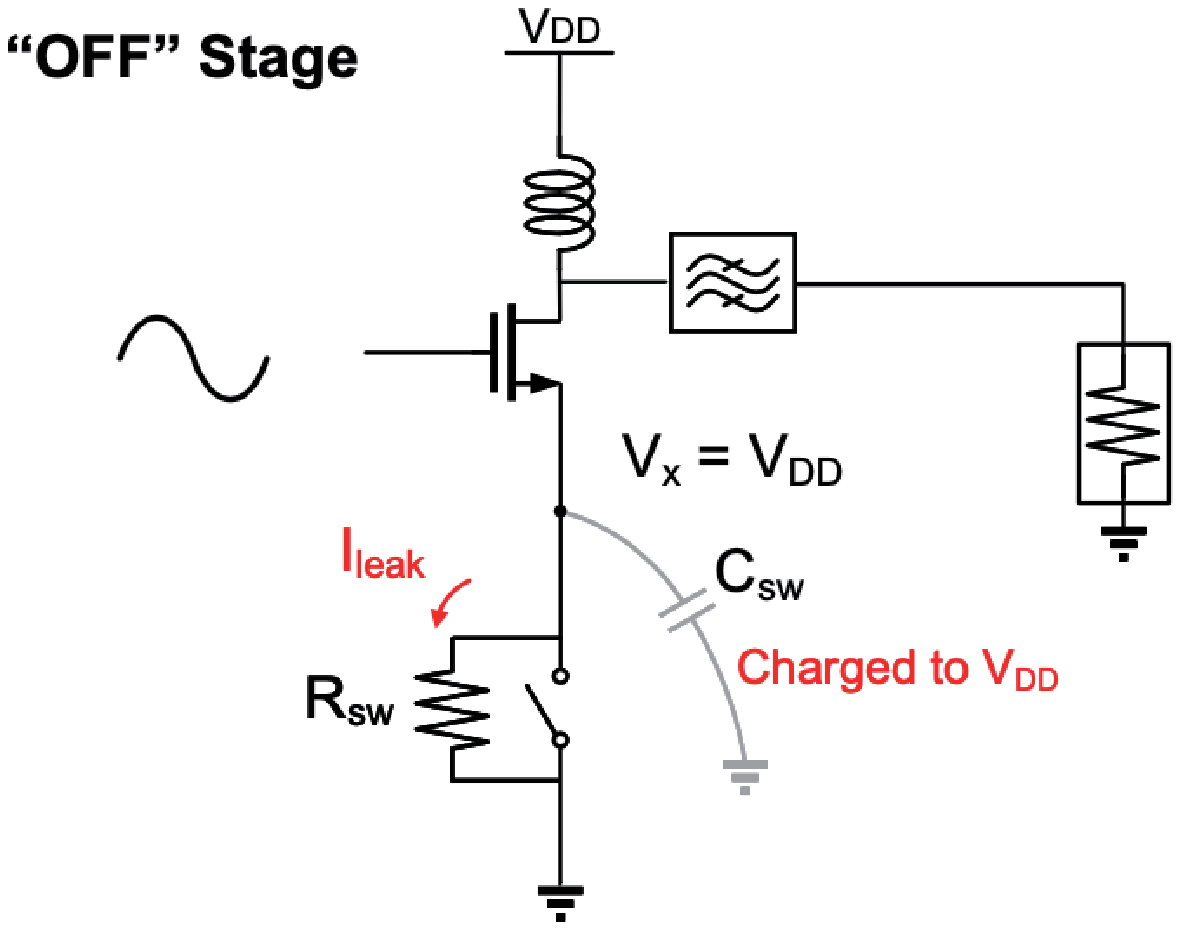}
}
\subfigure[]{
\includegraphics[width=0.31\linewidth]{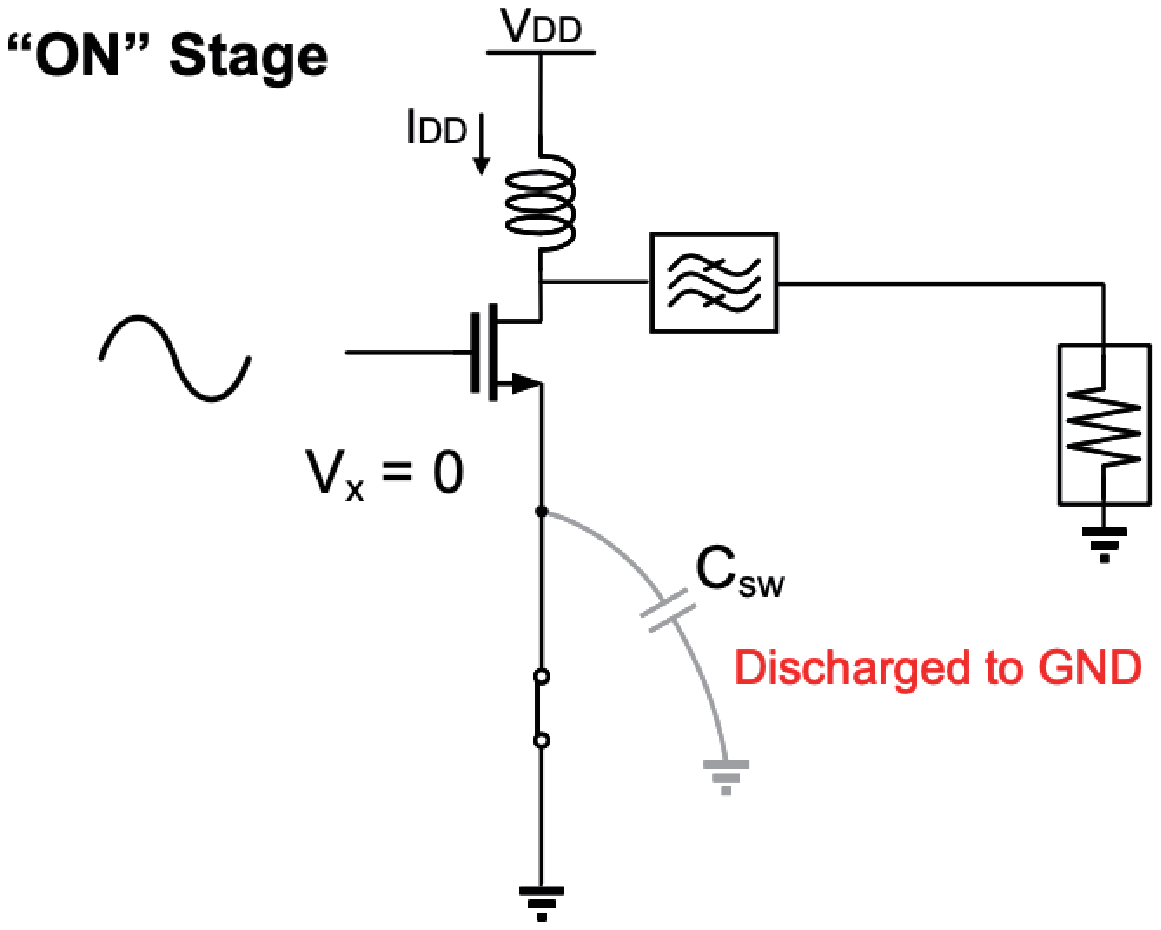}
}
\caption{(a) Schematic of a time-modulated PA cell with the upper NMOS biased as a typical class-A/B amplifier and the lower NMOS as a voltage-controlled current source to realize pulse modulation. 
(b) Efficiency and power model during the "OFF" state where the leakage current goes through the $R_{sw}$ and the switching capacitor $C_{sw}$ is charged to $V_{DD}$. 
(c) Efficiency and power model during the "ON" state where the switching unit is modeled as a short circuit and the switching capacitor $C_{sw}$ is discharged to GND.}
\label{fig:PA model}
\end{figure*}

When the STHS array is implemented in the power back-off (PBO) region, the output power is reduced by shortening the duration of the pulses. The relative pulse duration for the STHS working in the PBO region is denoted by the duty cycle ratio $\alpha$ where
\begin{equation}
\tau = \alpha \tau_{0}=\alpha \frac{T_p}{3}, \quad \alpha\in[0,1].
\label{Eq:back-offtau}
\end{equation}
Ideally, a duty cycle ratio of $10log(\alpha) = -p$ corresponds to a power back-off of $p$-dB.
While the duty cycle $\tau$ is designed to eliminate harmonics of $m = 3k$ for $k\in \mathbb{Z}$,
a change in the duty cycle in the PBO region will 
result in a decrease of harmonic efficiency $\zeta_{harm}$. To eliminate those harmonics, a systematic method is proposed for harmonic elimination in the back-off region.
Generally, to make $|A_{mn}|=0$
for a specific $m^{th}$ harmonic, according to   (\ref{Eq:AmnSimplified}), $t_{1n}$ and $s_{1n}$
should satisfy
\begin{equation}
e^{-j m \pi \frac{2 t_{1 n}+\tau}{T_{p}}} e^{j \pi \frac{1-m}{2}}+
e^{-j m \pi \frac{2 s_{1 n}+\tau}{T_{p}}} e^{j \pi \frac{-1-m}{2}}=0,
\label{Eq:eliminateM1}
\end{equation}
which means there exists a $k$ belongs to $\mathbb{Z}$, such that
\begin{equation}
\begin{aligned}
-m\left(\frac{2 t_{1 n}}{T_{p}}+\frac{\alpha}{3}\right)&+\frac{1-m}{2}=\\
-m&\left(\frac{2 s_{1n}}{T_{p}}+\frac{\alpha}{3}\right)+\frac{-1-m}{2}+2 k+1.
\end{aligned}
\label{Eq:eliminateM2}
\end{equation}
To simplify, If we let $k^{\prime} = -k$,   (\ref{Eq:eliminateM2}) yields
\begin{equation}
    \frac{t_{1n}}{T p}=\frac{s_{1n}}{T_{p}}+\frac{k^{\prime}}{|m|}.
    \label{Eq:eliminateM3}
\end{equation}

To optimize the harmonic efficiency in the power back-off region, since the power on each harmonic are related to the magnitude of the corresponding Fourier coefficients,
$P_{(-3)}$ is of our major  consideration 
as it contains the highest unsuppressed harmonic power.
Therefore, to ensure $|A_{-3n}|=0$, 
the relationship between  $t_{1n}$ and $s_{1n}$
should satisfy
\begin{equation}
    \frac{s_{1n}}{T p} = \frac{t_{1n}}{T_{p}}-\frac{1}{3}.
    \label{Eq:eliminate3}
\end{equation}
By substituting   (\ref{Eq:eliminate3}) into   (\ref{Eq:AmnSimplified}), the $A_{m n}$
now becomes
\begin{equation}
\begin{aligned}
A_{mn}&=\frac{4}{m \pi} \sin \left(m \pi \frac{\tau}{T_{p}}\right) \sin \left(\frac{m \pi}{2}\right) \operatorname{sgn}(m)\\
 \cdot&\left[e^{-jm\pi \frac{2 t_{1n}+\tau}{T_{p}}} e^{j \pi \frac{1-m}{2}}+e^{-j m\pi \frac{2 t_{1n}+\tau}{T_p}} e^{j\pi\left(\frac{-1-m}{2}+\frac{2}{3} m\right)}\right]
\end{aligned}.
\label{Eq:finalAmn}
\end{equation}
For an STHS array to achieve beamforming in the first harmonic,
$A_{1n}$
in the peakmode is considered
\begin{equation}
A_{1n}=\frac{4}{\pi} \frac{\sqrt{3}}{2}  e^{-j \pi \frac{2 t_{1n}+\tau}{\tau_{p}}} \cdot \left(1+e^{-j \frac{\pi}{3}}\right).
\end{equation}
Therefore, the array factor of the first harmonic is
\begin{equation}
\begin{aligned}
A&F_{1}(\theta)\\&=\sum_{n=1}^{N}I_n \frac{4}{\pi} \frac{\sqrt{3}}{2} e^{-j \pi\left(\frac{2 t_{1n}+\tau}{T_{p}}\right)} \cdot \left(1+e^{-j \frac{\pi}{3}}\right) \cdot e^{j \beta(n-1) d \sin \theta}\\
&=\sum_{n=1}^{N} I_n \frac{6}{\pi} e^{-j \pi\left(\frac{2t_{1n}}{T p}+\frac{1}{2}\right)} e^{j \beta(n-1) d \sin \theta}.
\end{aligned}
\end{equation}
To ensure the power of the first harmonic at  given direction $\theta$ is maximized,
$t_{1n}$  for each antenna element should
satisfy
\begin{equation}
\frac{t_{1 n}}{T_{p}}=\frac{1}{2}\left[\frac{(n-1) \beta d \sin \theta}{\pi}-\frac{1}{2}\right].
\end{equation}

The conceptual illustrations of the STHS transmitter array operating in the peakmode and the PBO mode are shown in Fig. \ref{fig:STHSPA}.
Compared with the pulse modulation back-off technique in Fig. \ref{fig:TimemodulatedPA}, it could be seen that all $m^{th}$ harmonics for $m \in \{2k\text{, } 3k\text{, }4k-1, \quad  k \in \mathbb{Z} \}$ are suppressed, and therefore, the array harmonic efficiency is improved. Moreover, in the power back-off region, the undesired $m = -3$ harmonic is suppressed leading to an improved back-off harmonic efficiency.

In order to demonstrate the efficiency enhancement of the STHS back-off technique, a simulation is performed in Matlab by choosing a carrier frequency $f_0 = 77$ GHz, and a pulse frequency $f_p = 1$ GHz.
In this simulation, 5 antenna elements are used and the spacing between each antenna is set to be 
half wavelength (1.95 mm). The mutual coupling between antennas is neglected.
The first harmonic ($m = 1$) targeted at $\theta= 20^{\circ}$ is simulated to demonstrate the change of harmonic power in the peakmode and the PBO mode with a duty cycle ratio of $10log(\alpha) = -6$, which ideally corresponds to a 6-dB PBO.
The radiation patterns of the first harmonic ($m = 1$) are shown together with the normalized relative power patterns at other harmonics $m = \{$-3, 5, -7$\}$ in Figs. \ref{fig:ideal0} and \ref{fig:ideal6}.  It is seen that the beamforming without phase shifters is achieved with the first harmonic targeted at  $\theta= 20^{\circ}$, and
due to the harmonic-suppression algorithm deduced in (\ref{Eq:eliminate3}), the $m = -3$ harmonic is suppressed in both the peakmode and the PBO region, otherwise it will consume much power indicated by the Fourier coefficients.

\begin{figure*}[h]
\begin{minipage}[h]{0.48\linewidth}
%
%
\definecolor{mycolor1}{rgb}{0,0,1}%
\begin{tikzpicture}
\pgfplotsset{
  grid style = {
    dash pattern = on 0.025mm off 0.95mm on 0.025mm off 0mm, 
    line cap = round,
    gray,
    line width = 1pt
  }
}
\begin{axis}[%
width=2.5in,
height=2.1in,
scale only axis,
line width=1 pt,
tick label style={font=\boldmath}, 
xmin=1,
xmax=9,
xlabel style={font=\bfseries\color{white!15!black}},
xlabel={\textbf{Switching / Amplifying Transistor Width Ratio}},
ymin=7.2,
ymax=9,
ylabel style={font=\bfseries\color{white!15!black}},
ylabel={\textbf{Switching Power Ratio}},
axis background/.style={fill=white},
xmajorgrids,
ymajorgrids
]
\addplot [color=mycolor1, line width=1.5pt, mark=diamond, mark options={solid, mycolor1}, forget plot]
  table[row sep=crcr]{%
1	7.33518251434488\\
2	8.16440167956139\\
3	8.59918485200716\\
4	8.65340905624584\\
5	8.66818802926048\\
6	8.66995813110648\\
7	8.65518519074774\\
8	8.63739107345217\\
9	8.61952374921452\\
};
\end{axis}
\node[below right, align=left]
at (rel axis cs:0.43,0.945) {\large $\textbf{SPR}_{\textbf{max}}$: \textbf{8.7}};

\node[below right, align=left]
at (rel axis cs:0.5825,0.831) {\small $\triangle$};
\end{tikzpicture}%
 \caption{Switching power ratio (SPR) vs. switching / amplifying transistor width ratio.}
 \label{fig:swPowerRatio}
\end{minipage}%
\hspace{5mm}
\begin{minipage}[h]{0.48\linewidth}
 \input{Figs/p1.tex}
 \caption{Comparison of the simulated transient output power  between a single-stage PA and a 2-stage PA chain.}
 \label{fig:p1}
\end{minipage}%
\end{figure*}

 \subsection{Circuit Efficiency Modeling and Optimization}
In the previous subsection, we focused on the array factor and optimize the harmonic efficiency using the pulse modulation technique. In this subsection, the performance of the STHS transmitter array is optimized in the circuit perspective.
As shown in Fig. \ref{fig:PA model} (a), the time-modulated PA structure consists of a classical common source amplifier with its source connected to another transistor that serves as a voltage-controlled current source. 
The upper NMOS is biased as a typical class-A/B amplifier and the lower NMOS is used as a switching device.
 During the "ON" state, the source of the upper NMOS is connected directly to the GND and it will function as an ordinary class-A/B  amplifier. During the "OFF" state, the output will remain at VDD since the upper NMOS is floating. By switching between the "ON" and the "OFF" state at a rate faster than the data bandwidth, this structure is able to realize the pulse modulation of a given RF signal. Compared with the existing structure involving SPST switches along with the signal path, this stacking time-modulated PA structure reduces the insertion loss when the switch is "ON" and allows smaller power leakage when the switch is "OFF".
\subsubsection{Time-modulated PA Cell Switching performance}
In terms of the circuit switching performance, the maximum output power when the switch is at the "ON" stage $P_{ON,max}$, and the "OFF" stage $P_{OFF,max}$ is examined. 
A smaller switching unit reduces the $P_{ON,max}$  because
the maximum current the current source can provided is small and the gain of the amplifier is therefore limited. 
For switching units with a large transistor size, there will be more leakage power during the "OFF" stage, and  $P_{OFF,max}$ will be larger.
Therefore, there exists a design trade-off between the output power and power leakage for the time-modulated PA.
In STHS transmitter array design, 
the size ratio between the switching and the amplifying unit is designed to optimized the switching power ratio (SPR) i.e.,
 \begin{equation}
SPR = \frac{P_{ON,max}}{P_{OFF,max}}.
\end{equation}
The simulation result for SPR under different transistor width ratios between the bottom switching NMOS and the top amplifying NMOS is shown in Fig. \ref{fig:swPowerRatio}, showing that the maximum SPR can be reached for a transistor width ratio of 6.

\begin{figure*}[h]
\subfigure[]{
%
%
\definecolor{mycolor2}{RGB}{146,205,220}%
\definecolor{mycolor4}{RGB}{215,227,191}%
\definecolor{mycolor3}{RGB}{229,185,181}%
\definecolor{mycolor1}{RGB}{254,229,153}%

\begin{tikzpicture}

\begin{axis}[%
width=2.3in,
height=2.3in,
at={(1.236in,0.481in)},
scale only axis,
xmin=-1.2,
xmax=1.2,
ymin=-1.2,
ymax=1.2,
axis line style={draw=none},
ticks=none,
axis x line*=bottom,
axis y line*=left,
legend style={at={(0.05,0.12)}, anchor=south west, legend cell align=left, align=left, draw=white!15!black}
]

\addplot[area legend, draw=black, fill=mycolor1]
table[row sep=crcr] {%
x	y\\
0	0\\
6.12323399573677e-17	1\\
-0.062282100536466	0.998058585431119\\
-0.124322370318208	0.992241879905531\\
-0.185879917578003	0.982572468696938\\
-0.246715724877701	0.969087896476925\\
-0.3065935771721	0.951840521535418\\
-0.365280978991603	0.93089731248239\\
-0.422550057182408	0.90633958822019\\
-0.478178445699022	0.878262702196141\\
-0.531950149013625	0.846775672161395\\
-0.583656380789801	0.812000756873632\\
-0.633096374564205	0.774072981387194\\
-0.68007816328844	0.733139612773872\\
-0.724419324704301	0.689359588310023\\
-0.765947689658243	0.642902898350287\\
-0.804502010604777	0.593949926284085\\
-0.839932587703147	0.542690748137735\\
-0.872101850076227	0.489324394541722\\
-0.900884889974733	0.434058077928764\\
-0.926169947772676	0.377106387963343\\
-0.947858845910886	0.318690458326701\\
-0.965867370103708	0.259037108092542\\
-0.980125596328676	0.198377961027319\\
-0.990578162329553	0.136948546234729\\
0	0\\
}--cycle;
\addlegendentry{\textbf{$P_{out,ON}$}}

\node[above left, align=right]
at (axis cs:-0.723,0.829) {23\%};

\addplot[area legend, draw=black, fill=mycolor2]
table[row sep=crcr] {%
x	y\\
-0	0\\
-0.990578162329553	0.136948546234729\\
-0.997205670488902	0.074705091826319\\
-0.999925955829357	0.0121689300576221\\
-0.998728359806475	-0.0504149116657815\\
-0.993617574806839	-0.112801219120075\\
-0.984613625762482	-0.174745552054579\\
-0.971751791689765	-0.236005201950575\\
-0.955082467460147	-0.296340142994898\\
-0.934670966344434	-0.355513972542235\\
-0.910597264104184	-0.413294837381227\\
-0.882955685632958	-0.469456342175109\\
-0.851854535375216	-0.52377843651746\\
-0.817415672970925	-0.576048277127438\\
-0.779774035788601	-0.626061061806241\\
-0.739077110217552	-0.673620831887252\\
-0.695484353790919	-0.718541240035691\\
-0.649166570403716	-0.760646280389432\\
-0.600305241073876	-0.799770978180151\\
-0.54909181286851	-0.835762036132758\\
-0.495726948781461	-0.868478435110408\\
-0.440419741501286	-0.897791986651663\\
-0.383386894150249	-0.923587835234866\\
-0.324851871204329	-0.945764908301765\\
-0.265044022921075	-0.964236312277137\\
-0.204197686705977	-0.97892967303271\\
-0.142551268938323	-0.989787419461408\\
-0.080346310854121	-0.996767009050828\\
-0.0178265421461188	-0.999841094572089\\
0.0447630740109259	-0.998997631230972\\
0.107177300767976	-0.994239923861485\\
0.169171588481736	-0.985586613976958\\
0.230503032902096	-0.9730716067294\\
0.290931326912632	-0.956743938063291\\
0.350219702095103	-0.936667582584354\\
0.408135856428707	-0.912921202896069\\
0.464452864489241	-0.885597841386111\\
0.51895006658182	-0.854804555670315\\
0.571413933323407	-0.820661999122582\\
0.621638902287546	-0.78330394813427\\
0.669428183433203	-0.742876777955349\\
0.714594530161892	-0.699538889171078\\
0.756960972981955	-0.653460087061338\\
0.796361512905378	-0.604820916274443\\
0.832641771860295	-0.553811953422232\\
0.865659597570742	-0.500633060368232\\
0.89528562053365	-0.445492601134606\\
0.92140376091075	-0.388606625496175\\
0.943911683349278	-0.33019802246036\\
0.962721197949434	-0.270495646949828\\
0.977758605807541	-0.209733423109657\\
0.988964987780977	-0.148149427752425\\
0.99629643534349	-0.0859849575324346\\
0.999724222626325	-0.0234835835040361\\
0.999234918971105	0.0391098032315217\\
0.99483044155346	0.101549951051426\\
0.986528047871196	0.163592208749214\\
0.974360268126464	0.224993484120153\\
0.958374777766847	0.285513196438179\\
0.938634210684785	0.344914219092445\\
0.915215913807238	0.40296380869004\\
0.888211644037168	0.459434516984512\\
0.857727208734256	0.514105082057105\\
0.823882051143522	0.566761295258897\\
0.786808782396213	0.617196840517017\\
0.746652661916645	0.665214102716402\\
0.703571028270872	0.710624941989702\\
0.657732682687205	0.753251430881543\\
0.609317227664048	0.792926551498813\\
0.558514363256519	0.829494849915396\\
0.505523143799097	0.862813045267327\\
0.450551197976609	0.892750591151775\\
0.393813915299435	0.919190187130242\\
0.335533602170477	0.942028238331794\\
-0	0\\
}--cycle;
\addlegendentry{\textbf{$P_{DC,ON}$}}

\node[below, align=center]
at (axis cs:0.374,-1.035) {72\%};

\addplot[area legend, draw=black, fill=mycolor3]
table[row sep=crcr] {%
x	y\\
0	0\\
0.335533602170477	0.942028238331794\\
0.301528287043517	0.953457231401914\\
0	0\\
}--cycle;
\addlegendentry{\textbf{$P_{leak}$}}

\node[above, align=center]
at (axis cs:0.45,0.95) {$\text{\textless{} 1\%}$};

\addplot[area legend, draw=black, fill=mycolor4]
table[row sep=crcr] {%
x	y\\
0	0\\
0.301528287043517	0.953457231401914\\
0.242591357164354	0.970128565412418\\
0.182744347002031	0.983160466881577\\
0.122211772320799	0.99250404669513\\
0.0612207207768323	0.998124252459364\\
-5.82016719913287e-16	1\\
0	0\\
}--cycle;
\addlegendentry{\textbf{$P_{dyn}$}}

\node[above, align=center]
at (axis cs:0.14,1) {5\%};
\end{axis}

\end{tikzpicture}%
}
\subfigure[]{
%
%
\definecolor{mycolor2}{RGB}{146,205,220}%
\definecolor{mycolor4}{RGB}{215,227,191}%
\definecolor{mycolor3}{RGB}{229,185,181}%
\definecolor{mycolor1}{RGB}{254,229,153}%
\begin{tikzpicture}

\begin{axis}[%
width=2.3in,
height=2.3in,
at={(1.236in,0.481in)},
scale only axis,
xmin=-1.2,
xmax=1.2,
ymin=-1.2,
ymax=1.2,
axis line style={draw=none},
ticks=none,
axis x line*=bottom,
axis y line*=left,
legend style={at={(0.05,0.12)}, anchor=south west, legend cell align=left, align=left, draw=white!15!black}
]

\addplot[area legend, draw=black, fill=mycolor1]
table[row sep=crcr] {%
x	y\\
0	0\\
6.12323399573677e-17	1\\
-0.0611741474252236	0.998127107981141\\
-0.122119149705501	0.992515447373594\\
-0.182606720024112	0.983186038246087\\
-0.242410285005676	0.970173826550411\\
-0.301305833411088	0.953527553221431\\
-0.359072755235238	0.933309571604071\\
-0.415494668064448	0.909595613891148\\
-0.470360227598266	0.882474507446933\\
-0.523463919299569	0.852047842079032\\
-0.574606828207638	0.818429589504899\\
-0.623597384030644	0.781745676438404\\
-0.670252078726587	0.74213351289555\\
-0.714396153884788	0.699741477486238\\
-0.75586425533316	0.654728361620029\\
-0.794501052519225	0.607262774707822\\
-0.830161820344814	0.557522512587416\\
-0.862712981275032	0.505693891538691\\
-0.89203260569086	0.45197105038307\\
-0.918010868611167	0.396555223281413\\
-0.940550461073379	0.339653985954315\\
0	0\\
}--cycle;
\addlegendentry{\textbf{$P_{out,ON}$}}

\node[above left, align=right]
at (axis cs:-0.632,0.9) {19\%};

\addplot[area legend, draw=black, fill=mycolor2]
table[row sep=crcr] {%
x	y\\
-0	0\\
-0.940550461073379	0.339653985954315\\
-0.959814773133317	0.28063428385541\\
-0.975379537742119	0.220532894040709\\
-0.987184761464751	0.159581473648361\\
-0.995184941787025	0.0980149562074883\\
-0.999349242502259	0.0360706461012388\\
-0.999661612567557	-0.0260126961084753\\
-0.99612084797157	-0.0879957739689842\\
-0.988740596375287	-0.149639677490433\\
-0.97754930450796	-0.210706804009752\\
-0.962590108520917	-0.270961774015614\\
-0.943920667721907	-0.330172338404399\\
-0.921612942330813	-0.388110273670179\\
-0.89575291611338	-0.444552261578293\\
-0.866440264962037	-0.499280749931854\\
-0.833787972701256	-0.552084791113403\\
-0.797921895598296	-0.602760855169628\\
-0.758980277257906	-0.651113614305147\\
-0.717113215770787	-0.696956695761567\\
-0.672482085169665	-0.740113400179904\\
-0.625258913422935	-0.780417382677482\\
-0.575625719363356	-0.817713294014118\\
-0.52377381110757	-0.851857379376296\\
-0.46990304867065	-0.882718032471314\\
-0.414221073617867	-0.910176302795707\\
-0.356942508722955	-0.934126354122697\\
-0.298288130717731	-0.954475872441479\\
-0.238484019321658	-0.971146421775927\\
-0.177760685831371	-0.984073746511287\\
-0.11635218462895	-0.993208019063515\\
-0.0544952110335965	-0.998514031936659\\
0.00757181102602399	-0.999971333428007\\
0.069609648002266	-0.99757430645792\\
0.131379178839606	-0.991332190220529\\
0.192642316652152	-0.98126904457182\\
0.253162926415015	-0.967423657292391\\
0.312707735132346	-0.949849394582319\\
0.371047230973857	-0.9286139953644\\
0.427956547914163	-0.903799310188602\\
0.483216332465148	-0.875500985744124\\
0.536613589160593	-0.843828096195064\\
0.587942501534201	-0.80890272276072\\
0.637005225426606	-0.770859483160971\\
0.683612651563596	-0.729845012740506\\
0.727585134466277	-0.686017399271833\\
0.768753184883604	-0.639545573615607\\
0.806958123078345	-0.590608658586948\\
0.84205269044845	-0.53939527853748\\
0.873901617126341	-0.486102832314281\\
0.902382143368354	-0.430936732398082\\
0.927384492724684	-0.374109613153393\\
0.948812295166006	-0.315840511242329\\
0.96658295853587	-0.256354021361172\\
0.980627986897128	-0.195879430553812\\
0.99089324454534	-0.134649834438845\\
0.997339164671544	-0.0729012387567348\\
0.999940901870105	-0.010871649700094\\
0.998688427903813	0.0511998434666567\\
0.993586570357115	0.113073989962259\\
0.984654994028482	0.174512299666156\\
0.971928125133643	0.235277962364096\\
0.955455018611843	0.295136760518649\\
0.935299169046569	0.35385797204641\\
-0	0\\
}--cycle;
\addlegendentry{\textbf{$P_{DC,ON}$}}

\node[below, align=center]
at (axis cs:0.374,-1.035) {61\%};

\addplot[area legend, draw=black, fill=mycolor3]
table[row sep=crcr] {%
x	y\\
0	0\\
0.935299169046569	0.35385797204641\\
0.914073690678667	0.405548132789538\\
0.889994114423726	0.455972012618239\\
0.863135626197731	0.504972168330345\\
0	0\\
}--cycle;
\addlegendentry{\textbf{$P_{leak}$}}

\node[above right, align=left]
at (axis cs:0.9,0.4) {3\%};

\addplot[area legend, draw=black, fill=mycolor4]
table[row sep=crcr] {%
x	y\\
0	0\\
0.863135626197731	0.504972168330345\\
0.830600428572794	0.556868860733559\\
0.794948979469957	0.606676289333672\\
0.756315036201564	0.654207586332829\\
0.714843545818767	0.699284423536842\\
0.670690101298343	0.741737681407934\\
0.624020357788075	0.781408083568402\\
0.575009411100801	0.818146794374646\\
0.523841140788946	0.851815977319594\\
0.470707520264146	0.882289312168508\\
0.415807896550301	0.909452468887965\\
0.359348242372261	0.933203536589938\\
0.301540383386169	0.953453405881652\\
0.242601203450738	0.970126103186722\\
0.182751830921147	0.983159075783247\\
0.122216809018397	0.992503426489481\\
0.0612232533867605	0.998124097116556\\
3.06161699786838e-16	1\\
0	0\\
}--cycle;
\addlegendentry{\textbf{$P_{dyn}$}}

\node[above right, align=left]
at (axis cs:0.547,0.954) {17\%};
\end{axis}

\end{tikzpicture}%
}
\subfigure[]{
%
%
\definecolor{mycolor2}{RGB}{146,205,220}%
\definecolor{mycolor4}{RGB}{215,227,191}%
\definecolor{mycolor3}{RGB}{229,185,181}%
\definecolor{mycolor1}{RGB}{254,229,153}%
\begin{tikzpicture}

\begin{axis}[%
width=2.3in,
height=2.3in,
at={(1.236in,0.481in)},
scale only axis,
xmin=-1.2,
xmax=1.2,
ymin=-1.2,
ymax=1.2,
axis line style={draw=none},
ticks=none,
axis x line*=bottom,
axis y line*=left,
legend style={at={(0.05,0.12)}, anchor=south west, legend cell align=left, align=left, draw=white!15!black}
]

\addplot[area legend, draw=black, fill=mycolor1]
table[row sep=crcr] {%
x	y\\
0	0\\
6.12323399573677e-17	1\\
-0.0617359380234735	0.998092517733883\\
-0.123236355633023	0.992377347912724\\
-0.184266630916751	0.982876293706685\\
-0.244593935539057	0.969625601300582\\
-0.303988122972483	0.952675821615972\\
-0.362222606498548	0.93209161746118\\
-0.419075223628048	0.907951516844981\\
-0.474329083643069	0.880347613395031\\
-0.527773395027385	0.849385215023942\\
-0.579204269628614	0.815182442183333\\
-0.628425500484293	0.777869777238495\\
-0.675249310344472	0.737589566682796\\
-0.719497068035272	0.694495478090856\\
-0.760999969930472	0.648751913882203\\
-0.799599683931355	0.600533384131871\\
-0.835148953498055	0.550023840820653\\
-0.867512159428028	0.497415976064821\\
0	0\\
}--cycle;
\addlegendentry{\textbf{$P_{out,ON}$}}
\node[above left, align=right]
at (axis cs:-0.551,0.952) {17\%};

\addplot[area legend, draw=black, fill=mycolor2]
table[row sep=crcr] {%
x	y\\
-0	0\\
-0.867512159428028	0.497415976064821\\
-0.896802726203281	0.442430638941701\\
-0.922612510495474	0.385728084895616\\
-0.944841336153896	0.327528394944499\\
-0.96340292589068	0.268057460976702\\
-0.978225236150955	0.207546108991253\\
-0.989250736737269	0.146229203187226\\
-0.996436634102964	0.0843447343795482\\
-0.999755037447817	0.022132896279396\\
-0.999193066971291	-0.0401648467755675\\
-0.994752903863222	-0.102306697022663\\
-0.986451781837912	-0.164051461769803\\
-0.974321920244489	-0.225159489542618\\
-0.958410399013157	-0.285393600249622\\
-0.938778975922704	-0.344520005755135\\
-0.915503846898516	-0.402309217286926\\
-0.888675350271454	-0.45853693615663\\
-0.858397616145472	-0.512984924335764\\
-0.824788162234885	-0.565441851508358\\
-0.787977437739966	-0.615704115312508\\
-0.748108317031262	-0.663576631587229\\
-0.705335545107781	-0.708873591557415\\
-0.659825136981425	-0.751419183018003\\
-0.611753733318837	-0.791048272718211\\
-0.561307914841633	-0.827607047297289\\
-0.508683478146067	-0.860953610284096\\
-0.454084675752875	-0.890958532843368\\
-0.397723423336946	-0.917505356131037\\
-0.339818477213802	-0.94049104330881\\
-0.280594585275323	-0.959826379463583\\
-0.220281614670231	-0.975436317879479\\
-0.159113659615069	-0.987260271318511\\
-0.0973281327985582	-0.99525234717932\\
-0.0351648439058896	-0.999381525621259\\
0.0271349311605057	-0.999631779962459\\
0.0893293867510532	-0.996002138884591\\
0.151177125995251	-0.988506690202862\\
0.212438097741282	-0.97717452618663\\
0.272874528272453	-0.962049630642872\\
0.332251844184175	-0.943190708200741\\
0.390339582839483	-0.920670956459852\\
0.446912286869346	-0.894577781886635\\
0.50175037924589	-0.865012460561468\\
0.554641015532127	-0.832089745093335\\
0.605378910000301	-0.795937419227697\\
0.65376713241243	-0.756695801876308\\
0.699617872370492	-0.714517202493954\\
0.74275316826953	-0.66956532991604\\
0.783005598024421	-0.622014656951441\\
0.820218928889325	-0.572049743196907\\
0.854248723847687	-0.519864518701361\\
0.884962902219183	-0.465661531260421\\
0.912242252307714	-0.409651160262667\\
0.935980894100694	-0.352050800138937\\
-0	0\\
}--cycle;
\addlegendentry{\textbf{$P_{DC,ON}$}}

\node[below, align=center]
at (axis cs:0.374,-1.035){53\%};
\addplot[area legend, draw=black, fill=mycolor3]
table[row sep=crcr] {%
x	y\\
0	-0\\
0.935980894100694	-0.352050800138937\\
0.95177627280519	-0.306792970136313\\
0.965384693145194	-0.260830201934053\\
0.976774886166499	-0.214268107179819\\
0	-0\\
}--cycle;
\addlegendentry{\textbf{$P_{leak}$}}

\node[right, align=left]
at (axis cs:0.95,-0.3) {2\%};

\addplot[area legend, draw=black, fill=mycolor4]
table[row sep=crcr] {%
x	y\\
0	-0\\
0.976774886166499	-0.214268107179819\\
0.988114622213903	-0.153718877718629\\
0.995704662076723	-0.0925863160444328\\
0.999516203080778	-0.031102407929033\\
0.999534781194835	0.0304995275667142\\
0.995760325918688	0.0919857234917907\\
0.988207160550697	0.153122852103563\\
0.976903947833751	0.213678910299618\\
0.961893581185927	0.273424100022862\\
0.943233021928598	0.33213170029891\\
0.920993083129677	0.389578927596569\\
0.895258160882274	0.445547781246511\\
0.866125914038502	0.499825870709959\\
0.833706893613783	0.552207221558045\\
0.798124123267986	0.602493057103324\\
0.759512632455389	0.650492552717312\\
0.718018944015071	0.696023559971559\\
0.673800518146213	0.738913297854285\\
0.627025154878328	0.778999008439554\\
0.577870357303907	0.816128574520863\\
0.526522657989895	0.850161096865327\\
0.473176911124135	0.880967428897926\\
0.418035553082921	0.90843066678677\\
0.361307834225689	0.93244659306962\\
0.303209024831995	0.95292407213819\\
0.243959598194122	0.969785396079443\\
0.183784393965293	0.982966579561487\\
0.122911764938424	0.992417602645037\\
0.0615727104932195	0.998102600599016\\
3.06161699786838e-16	1\\
0	-0\\
}--cycle;
\addlegendentry{\textbf{$P_{dyn}$}}

\node[above right, align=left]
at (axis cs:0.857,0.689) {28\%};
\end{axis}
\end{tikzpicture}%
}
\caption{Power distribution  for  
(a) $f_p = $ 1 GHz in the peakmode,
(b)  $f_p = $   1 GHz with $10log(\alpha) = -6$,
(c)  $f_p = $   2 GHz with $10log(\alpha) = -6$.
We denote $P_{out,ON}$ the output power during the "ON" stage, $P_{DC,ON}$ the power delivered from the DC source during the "ON" stage, $P_{leak}$ the power leakage during the "OFF" stage, and $P_{dyn}$ the dynamic power loss resulting from each switching.}
\label{fig:powerchart}
\end{figure*}
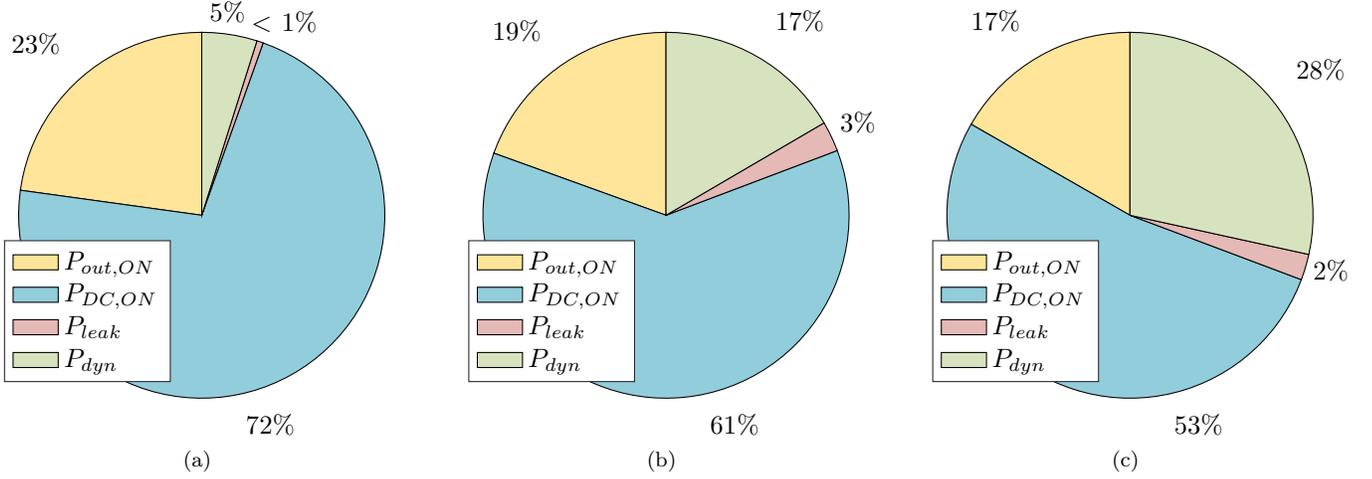

Moreover, a 2-stage cascaded power amplifying chain is used to further improve the SPR.
The simulated transient response at the output stage is shown in Fig. \ref{fig:p1}. The SPR increases from 8.7 for an individual time-modulated power amplifier to 809 for a 2-stage cascaded time-modulated power amplifier. This 92 times improvement is because
a 2-stage cascaded time-modulated power amplifier will amplify the $P_{ON, max}$ twice while blocking the $P_{OFF,max}$ twice.

\subsubsection{Time-modulated PA Cell Efficiency Model}
While the theoretical value for $\zeta_{harm}$ can be reached by combining   (\ref{Eq:tmaEff}) and   (\ref{Eq:harmonicsPower}) given in the previous section, the remaining $\zeta_{circ}$ part will be analyzed in the following according to the power and efficiency model of a single time-modulated PA cell established (Fig. \ref{fig:PA model}).  Three mechanisms are analyzed separately:
\begin{enumerate}
    \item "ON" Stage Efficiency:
When the STHS PA cell is at the "ON" stage, the drain-source voltage $v_{D S}$ and drain current $i_{D}$ can be written as
\begin{equation}
\left\{
\begin{aligned}
i_{D}&=I_{D D}-\frac{v_{pk}}{R_{L}} \cos \left(2\pi f_c t\right),\\
v_{D S}&=V_{D D}+v_{p k}\cos \left(2\pi f_c t\right)
\end{aligned}
\right.
\end{equation}
where $V_{D D}$ is the system voltage, $I_{D D}$ is the bias current, and $v_{pk}$ is the peak voltage of the load.

The output power during the "ON" stage can be expressed as 
\begin{equation}
P_{out,ON}=\frac{v_{p k}^{2}}{2R_{L}}.
\end{equation}
The power delivered from the DC source during the "ON" stage is
\begin{equation}
P_{DC,ON}=V_{DD}I_{DD}.
\end{equation}

\item "OFF" Stage Power Leakage:
When STHS PA cell is at the "OFF" stage, the system contributes neglegible output power, i.e.,
\begin{equation}
P_{out,OFF}=0.
\end{equation}
In this stage, the DC power comes from the leakage current of the switching NMOS. To model subthreshold leakage current in the switching unit, an equivalent switching resistor $R_{sw}$ is used. Then the DC power during the "OFF" stage is 
\begin{equation}
P_{DC,OFF}=P_{leak}=V_{DD}^2/R_{sw}.
\end{equation}

\item Dynamic Switching Loss: The dynamic power should be considered to model the power consumption for each switching period. During the "OFF" stage, since the lower NMOS is in the cutoff region, the source voltage of the upper NMOS ($V_{X}$) will be equal to the output bias drain voltage, i.e.,
\begin{equation}
V_{X,OFF}=V_{DD}.
\end{equation}
During the "ON" stage, the lower NMOS will be turned on with $V_{X}$ shorted to the ground and therefore,
\begin{equation}
V_{X,ON}=0.
\end{equation}
To model the power loss resulting from this ON-OFF stage voltage difference, a switching capacitor $C_{sw}$ is used.
For each rectangular pulse, the switching capacitor will go through a period of charging and discharging, leading to a dynamic power loss. In terms of the STHS array, there exist two pulses in one modulation period when the STHS technique is adopted. Consequently, the dynamic power can be expressed as
\begin{equation}
P_{dyn}=f_{p}V_{DD}^2C_{sw}.
\end{equation}
\end{enumerate}

 \begin{figure}[h]
\centering
%
%
\definecolor{mycolor1}{rgb}{0,0,1}%
\definecolor{mycolor2}{rgb}{1,0,1}%
\definecolor{mycolor3}{rgb}{1,0,0}%
\definecolor{mycolor4}{rgb}{0.5,0,1}%
\definecolor{mycolor5}{rgb}{0.45,0.65,0.2}%
\definecolor{mycolor6}{RGB}{255,140,0}%
\begin{tikzpicture}
\pgfplotsset{
  grid style = {
    dash pattern = on 0.025mm off 0.95mm on 0.025mm off 0mm, 
    line cap = round,
    gray,
    line width = 1pt
  }
}
\begin{axis}[%
width=2.4in,
height=2.2in,
scale only axis,
line width=1 pt,
tick label style={font=\boldmath}, 
xmin=-10.2,
xmax=0,
xlabel style={font=\bfseries\color{white!15!black}},
xlabel={\textbf{10log($\alpha$) }},
ymin=0,
ymax=35,
ylabel style={font=\bfseries\color{white!15!black}},
ylabel={\textbf{Circuit Drain Efficiency (\%)}},
axis background/.style={fill=white},
xmajorgrids,
ymajorgrids,
legend style={at={(0.98,0.02)},anchor=south east,legend cell align=left, align=left, draw=white!15!black}
]
\addplot [color=mycolor1, line width=1.5pt, mark=o, mark options={solid, mycolor1}]
  table[row sep=crcr]{%
-9.704	27.6\\
-8.758	28.1\\
-7.828	28.3\\
-6.854	28.6\\
-5.901	28.8\\
-4.919	29\\
-3.943	29.1\\
-2.968	29.2\\
-1.977	29.3\\
-0.989000000000001	29.4\\
0	29.4\\
};
\addlegendentry{\textbf{$f_p$ = 200 MHz}}

\addplot [color=mycolor2, line width=1.5pt, mark=diamond, mark options={solid, mycolor2}]
  table[row sep=crcr]{%
-10.183	22.5\\
-8.984	24.3\\
-8.087	24.8\\
-7.07	25.7\\
-6.035	26.3\\
-5.039	27\\
-3.994	27.7\\
-3.003	27.9\\
-2.003	28.3\\
-1.002	28.5\\
0	28.8\\
};
\addlegendentry{\textbf{$f_p$ = 500 MHz}}

\addplot [color=mycolor3, line width=1.5pt, mark=triangle, mark options={solid, mycolor3}]
  table[row sep=crcr]{%
-9.824	20.4\\
-9.149	20.4\\
-7.988	22.5\\
-7.007	23.5\\
-6.02	24.5\\
-4.993	25.5\\
-4.031	26\\
-3.016	26.7\\
-2.006	27.3\\
-1.013	27.7\\
0	28.1\\
};
\addlegendentry{\textbf{$f_p$  =  1 GHz}}

\addplot [color=mycolor4, line width=1.5pt, mark=square, mark options={solid, mycolor4}]
  table[row sep=crcr]{%
-10.039	15.3\\
-9.249	16.3\\
-8.023	18.2\\
-6.976	19.8\\
-6.158	20.6\\
-5.067	22.2\\
-4.064	23.1\\
-3.045	24.4\\
-2.019	25.3\\
-1.016	26\\
0	26.8\\
};
\addlegendentry{\textbf{$f_p$ = 2 GHz}}

\addplot [color=mycolor5, dashed, line width=1.5pt]
  table[row sep=crcr]{%
-10	25.4133902786616\\
-9	26.5264551042083\\
-8	27.4825816578236\\
-7	28.2926272076369\\
-6	28.9709151991144\\
-5	29.5333254163348\\
-4	29.9958687558852\\
-3	30.373735161766\\
-2	30.6807381781792\\
-1	30.9290578534496\\
0	31.1291882810883\\
};
\addlegendentry{\textbf{200 MHz Model}}

\addplot [color=mycolor6, dashdotted, line width=1.5pt]
  table[row sep=crcr]{%
-10	12.8514740114077\\
-9	14.6519579788163\\
-8	16.4866764401325\\
-7	18.3076627000784\\
-6	20.0683632841256\\
-5	21.7282454734318\\
-4	23.2561760318899\\
-3	24.6320528906924\\
-2	25.8466889690823\\
-1	26.9003565168049\\
0	27.8005847267543\\
};
\addlegendentry{\textbf{2 GHz Model}}

\end{axis}

\end{tikzpicture}%
\caption{ Circuit drain efficiency for a 2-stage cascaded transmitting path vs. duty cycle ratio at different pulse frequencies.}
\label{fig:effCalculation}
\end{figure}
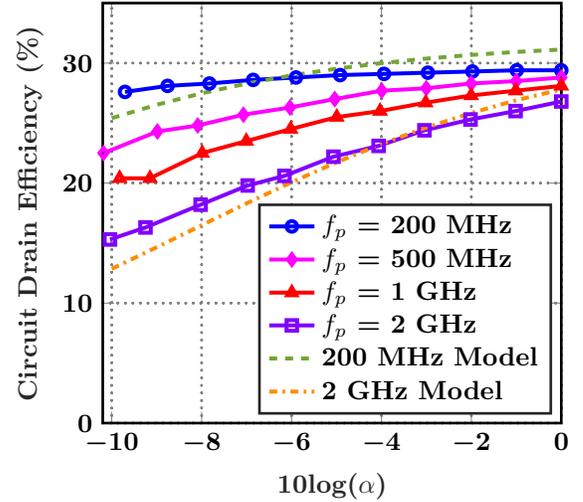

Hence, for an STHS PA in the PBO region with a duty cycle of $2\tau/T_{p}$, the overall drain efficiency of the circuit is
\begin{equation}
\begin{aligned}
\zeta_{circ} &= \frac{\frac{2\tau}{T_{p}}P_{out}}{\frac{2\tau}{T_{p}}P_{DC}+(1-\frac{2\tau}{T_{p}}) P_{leak}+P_{dyn}}\\
 &= \frac{\frac{2\tau}{T{p}}\frac{v_{p k}^{2}}{2R_{L}}}{\frac{2\tau}{T_{p}}V_{DD}I_{DD}+(1-\frac{2\tau}{T_{p}}) \frac{V_{DD}^2}{R_{sw}}+f_{p}V_{DD}^2C_{sw}}.
\end{aligned}
\label{Eq:circuitEff}
\end{equation}
  (\ref{Eq:circuitEff}) indicates that both an increase of pulse modulation frequency $f_p$
and a decrease of pulse duty cycle $\tau$ degrades the circuit efficiency. Intuitively, faster modulation frequency will lead to an increase in dynamic power and a smaller pulse duty cycle will result in more leakage power.

 \begin{figure*}[h]
  \begin{minipage}[h]{0.48\linewidth}
     \input{Figs/PBO0_8path.tex}
     \caption{Ideal radiation patterns for an 8-path   STHS array 
     with scanning angle $\theta=20^{\circ}$ in the peakmode. The $m = 5$ harmonic is suppressed.}
     \label{fig:8path0}
  \end{minipage}%
      \hspace{5mm}
    \begin{minipage}[h]{0.48\linewidth}
\input{Figs/PBO6_8path.tex}
     \caption{Ideal radiation patterns for an 8-path STHS array
     with scanning angle
    $\theta=20^{\circ}$ at $10log(\alpha)= -6$. Both $m = -3$ and $m = 5$ harmonics are suppressed.
    Ideally, this corresponds to a 6-dB PBO.}
     \label{fig:8path6}
  \end{minipage}%
 \end{figure*}
 
 \begin{figure*}[h]
    \centering
\subfigure[]{
\includegraphics[width=0.4\linewidth]{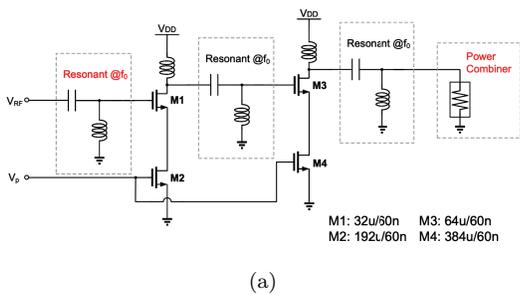}
}
\subfigure[]{
\includegraphics[width=0.26\linewidth]{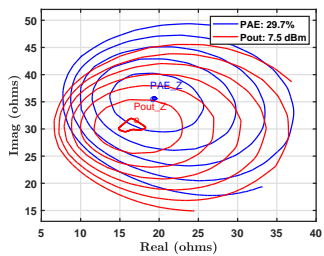}
}
\subfigure[]{
\includegraphics[width=0.26\linewidth]{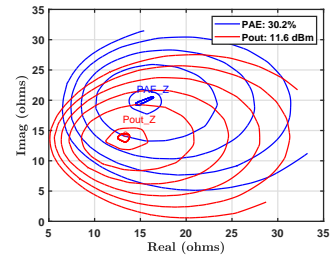}
}
\caption{(a) A 2-stage cascaded transmitting path in the STHS array.
(b) Power and efficiency contours for \textbf{M1}.
(c) Power and efficiency contours for \textbf{M3}.}
\label{fig:2stage}
\end{figure*}

According to the simulation, the model parameter is set to be $C_{sw}/W \approx 0.35 \text{ }nF/m$ and $R_{sw}\cdot W \approx 5.4 \text{ }\Omega\cdot m$, where W is the width of the transistor. Figs. \ref{fig:powerchart1Gpeakmode}, \ref{fig:powerchart1GPbo6} and \ref{fig:powerchart2GPbo6} show the distribution of the power with different modulation frequencies and operation regions. It can be seen that at peak power mode with modulation pulse frequency at 1 GHz, the dynamic power and leak power counts for 5\% and <1\% of the total power consumption, respectively. As predicted by our model, when the circuit goes into the PBO region, the simulation at a duty cycle ratio of $10log(\alpha) = -6$ (i.e., $\tau \approx 0.25 \times \frac{T_p}{3}$) shows that both leak power and dynamic power begin to increase causing degeneration of circuit efficiency. Additionally, if the circuit runs at 2 GHz with a duty cycle ratio of $10log(\alpha) = -6$, the dynamic power and leak power counts for 28\% and 2\%, respectively, which suggests that a doubling of switching frequency causes the dynamic power loss to increase by 64\% while the leak power remains nearly the same.

The drain efficiency predicted by the model is shown in
Fig. \ref{fig:effCalculation}  together with  the simulated results of the 2-stage cascaded circuit drain efficiency $\zeta_{circ}$  in different PBO regions with different modulation pulse frequencies. The theoretical efficiencies predicted by the circuit efficiency model at 200 MHz and 2 GHz are shown in dashed lines. From the simulation, it is seen that the circuit power and efficiency model matches the simulation results well.

\subsection{Further Optimization for STHS harmonic efficiency}
According to the array harmonic elimination technique for $m = -3$ harmonic discussed in   (\ref{Eq:eliminateM1}), (\ref{Eq:eliminateM2}), and (\ref{Eq:eliminateM3}), it is seen that a further improvement in harmonic efficiency could be achieved by
adding four additional paths with phases $45^{\circ}$, $135^{\circ}$, $225^{\circ}$, and $315^{\circ}$ to eliminate $m = 5$ harmonic.

For the implementation where the pulse switch-ON time for the $45^{\circ}$ path of $n^{th}$ is $r_{1n}$,
to eliminate $m = 5$ harmonic, according to  (\ref{Eq:finalAmn}), the requirement for  $r_{1n}$ is
\begin{equation}
\begin{aligned}
\left[e^{j \pi \frac{1-m}{2}}\right.&+\left.e^{j \pi\left(\frac{-1-m}{2}+\frac{2}{3}m\right)}\right] \times\\&\left[e^{-j m \pi \frac{2t_{1n}+\tau}{T_p}} + 
 e^{-j m \pi \frac{2 r_{1n}+\tau}{T_p}}\cdot e^{-j\frac{\pi}{4}}\right] = 0.
\end{aligned}
\end{equation}
Then the relation between $t_{1n}$ and $r_{1n}$
can be derived as
\begin{equation}
-m \frac{2 t_{1 n}+\tau}{T p}=-m \frac{2 r_{1n} +\tau}{T_{p}}-\frac{1}{4}+2k+1, \quad\{k \in \mathbb{Z}\}.
\end{equation}
Take $m = 5$ and $k = 0$, we arrive at
\begin{equation}
r_{1n} = t_{1n} + \frac{3}{40}.
\end{equation}

\begin{figure*}[h]
  \begin{minipage}[h]{0.48\linewidth}
    \input{Figs/PBO0_theta20_ADS.tex}
    \caption{Circuit-generated radiation pattern for the 4-path STHS array with scanning angle
    $\theta =20^{\circ}$ in the peakmode.}
    \label{fig:simulated0}
  \end{minipage}
  \hspace{5mm}
  \begin{minipage}[h]{0.48\linewidth}
    \input{Figs/PBO6_theta20_ADS.tex}
    \caption{Circuit-generated radiation pattern for the 4-path  STHS array with scanning angle
    $\theta=20^{\circ}$ at $10log(\alpha)=  -6$. The $m=-3$ harmonic is suppressed. Ideally, this corresponds to a 6-dB PBO.}
    \label{fig:simulated3}
  \end{minipage}
  \end{figure*}%

\begin{figure*}[h]
  \begin{minipage}[h]{0.48\linewidth}
%
%

\definecolor{mycolor1}{RGB}{169,169,169}%
\definecolor{mycolor2}{RGB}{237, 124, 50}%
\definecolor{mycolor3}{rgb}{0,0,1}%
\begin{tikzpicture}
\pgfplotsset{
  grid style = {
    dash pattern = on 0.025mm off 0.95mm on 0.025mm off 0mm, 
    line cap = round,
    gray,
    line width = 1pt
  }
}
\begin{axis}[%
width=2.3in,
height=2.1in,
at={(0.28\linewidth,0.642in)},
scale only axis,
line width=1 pt,
tick label style={font=\boldmath}, 
xmin=-10,
xmax=0,
xlabel style={font=\bfseries\color{white!15!black}},
xlabel={\textbf{10log($\alpha$)}},
ymin=0,
ymax=100,
ylabel style={font=\bfseries\color{white!15!black}},
ylabel={\textbf{Harmonic Efficiency (\%)}},
axis background/.style={fill=white},
xmajorgrids,
ymajorgrids,
legend style={at={(0.98,0.02)}, anchor=south east, legend cell align=left, align=left, draw=white!15!black}
]
\addplot [color=mycolor1, dashed, line width=1.5pt]
  table[row sep=crcr]{%
-10	18.427608854719\\
-9	23.5499185715176\\
-8	29.6690835208647\\
-7	37.4259650571107\\
-6	47.8577792550172\\
-5	59.5383501458041\\
-4	72.7225419812046\\
-3	88.0815776276656\\
-2	87.8753140720473\\
-1	89.8746671997795\\
0	89.6619580533014\\
};
\addlegendentry{\textbf{Ideal 4-path}}
\addplot [color=mycolor2, dotted, line width=1.5pt]
  table[row sep=crcr]{%
-10	34.4297697727671\\
-9	43.2639198144861\\
-8	56.4537245331042\\
-7	71.0009010624349\\
-6	82.6874981133546\\
-5	84.9724935150279\\
-4	88.3624802760876\\
-3	95.1425266402301\\
-2	93.6509257490339\\
-1	95.5543551179582\\
0	95.3120334855854\\
};
\addlegendentry{\textbf{Ideal 8-path}}
\addplot [color=mycolor3, line width=1.5pt, mark=diamond, mark options={solid, mycolor3}]
  table[row sep=crcr]{%
-10	12.88\\
-9	18.62\\
-8	24.18\\
-7	31.66\\
-6	41.32\\
-5	52.41\\
-4	65.7\\
-3	82.08\\
-2	84.63\\
-1	84.88\\
0	86.28\\
};
\addlegendentry{\textbf{Simulated 4-path}}

\end{axis}
\end{tikzpicture}%
    \caption{Simulated STHS array harmonic efficiency.}
    \label{fig:TMAeff}
  \end{minipage}%
  \hspace{5mm}
  \begin{minipage}[h]{0.48\linewidth}
%
%

\definecolor{mycolor1}{rgb}{0,0,1}%
\definecolor{mycolor2}{rgb}{1,0,1}%
\definecolor{mycolor3}{rgb}{0,0,1}%

\begin{tikzpicture}
\pgfplotsset{
  grid style = {
    dash pattern = on 0.025mm off 0.95mm on 0.025mm off 0mm, 
    line cap = round,
    gray,
    line width = 1pt
  }
}
\begin{axis}[%
width=2.3in,
height=2.1in,
at={(0.28\linewidth,0.642in)},
scale only axis,
line width=1 pt,
tick label style={font=\boldmath}, 
xmin=-10,
xmax=0,
xlabel style={font=\bfseries\color{white!15!black}},
xlabel={\textbf{10log($\alpha$)}},
ymin=-20,
ymax=0,
ylabel style={font=\bfseries\color{white!15!black}},
ylabel={\textbf{Power Backoff (dB)}},
axis background/.style={fill=white},
xmajorgrids,
ymajorgrids,
legend style={at={(0.98,0.02)},anchor=south east,legend cell align=left, align=left, draw=white!15!black}
]

\addplot [color=mycolor3, line width=1.5pt, mark=diamond,, mark options={solid, mycolor3}]
  table[row sep=crcr]{%
-10 -18.0839427340671\\
-9  -15.8083045978518\\
-8  -13.5125383990575\\
-7  -11.3609922590417\\
-6  -9.21749823582924\\
-5  -7.15795976727914\\
-4  -5.21444766870727\\
-3  -3.23272788662766\\
-2  -2.08985795555479\\
-1  -1.08404765537223\\
0   0\\
};

\end{axis}
\end{tikzpicture}%
    \caption{Simulated relationship between the duty cycle ratio $\alpha$ and the output power. Since in  $10log(\alpha) < -3$ region the harmonic efficiency begins to degrade faster, the slope of the output power is steeper.}
    \label{fig:aVsPower}
  \end{minipage}
\end{figure*}%

A similar simulation is performed with a carrier frequency of $f_0 = 77$ GHz,
a pulse frequency of $f_p = 1$ GHz 
with 5 antenna elements spaced at
half wavelength (1.95 mm). The result is shown in Figs. \ref{fig:8path0} and \ref{fig:8path6}. Compared with Figs. \ref{fig:ideal0} and \ref{fig:ideal6}, it can be seen that both $m = -3$ and $m = 5$ harmonics are eliminated in the peakmode results in harmonic efficiency improvement from 89.7\% to 95.3\% in the peakmode and 47.8\% to 82.7\% in the PBO mode with duty cycle ratio $10log(\alpha) = -6$.

\section{Circuit Simulation}
\subsection{Simulation Setup}
In this section, circuit simulation is performed to validate the effectiveness of the STHS array. A 65-nm NMOS process is used in the simulation. The circuit structure of one transmitter path is shown in Fig. \ref{fig:2stage}(a). Design challenges and considerations of major blocks are discussed in this section. The DC voltage is set to be 1.2 V, and the bias base voltage is set to be 0.7 V for a class-A/B operation. The carrier frequency is $f_0 = 77$ GHz, and the pulse frequency $f_p$ is chosen to be 1 GHz.
The spacing between each antenna is set to be 
half wavelength (1.95 mm).

To optimize the tradeoffs between PA maximum output power and efficiency, we design the L-type matching networks according to the load-pull simulation results shown in Figs. \ref{fig:2stage}(b) and \ref{fig:2stage}(c), and the matching network output load impedance for M1 and M2 is designed  to be $18+33j$
and $ 14+15j$, respectively.

\begin{figure*}[h]
  \begin{minipage}[h]{0.48\textwidth}
    \includegraphics[width=1\linewidth]{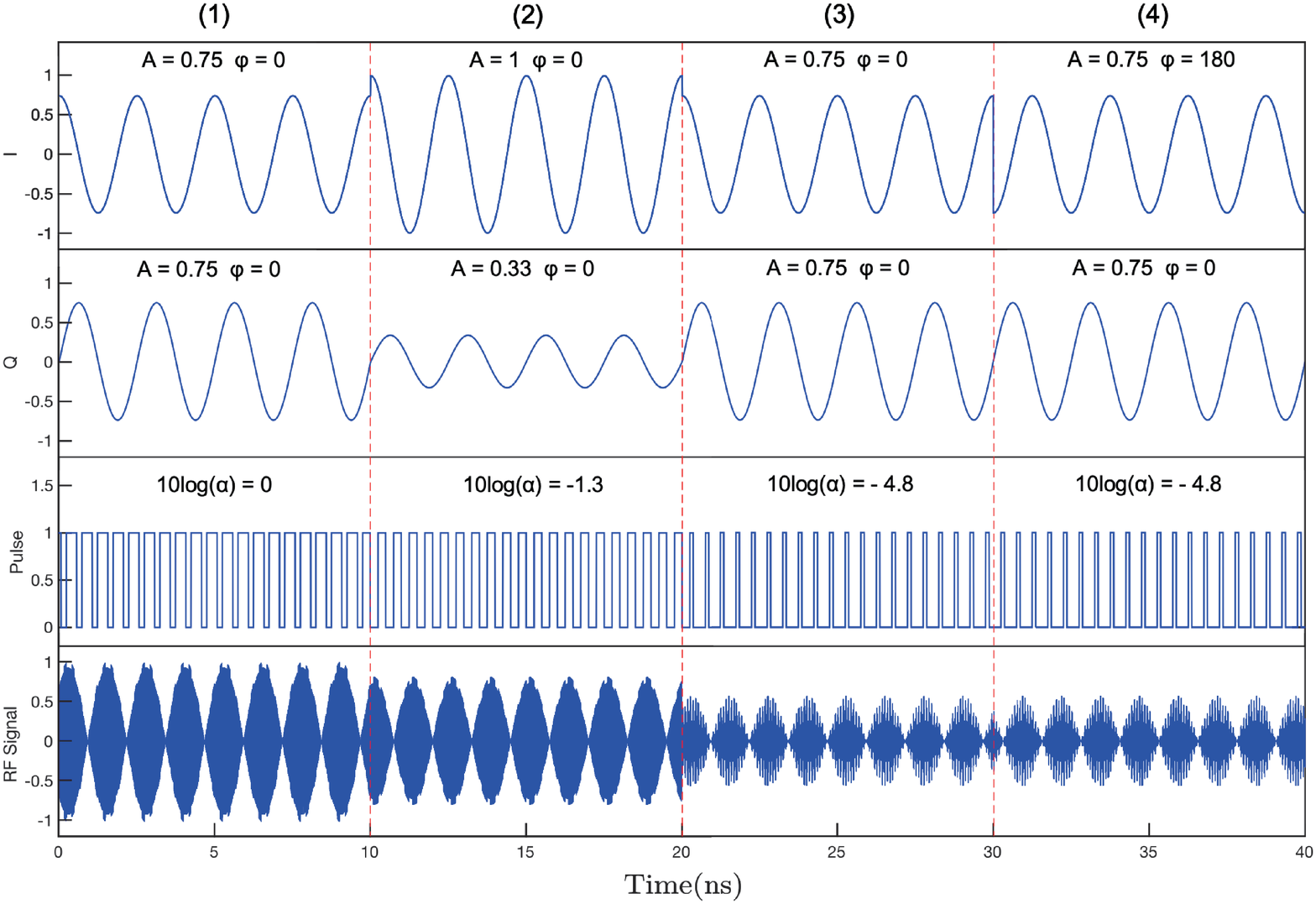}
    \caption{Illustration of the transient normalized amplitude and phase for a 4-bit symbol sequence transmitted through a 16 QAM.}
    \label{fig:qamsignal}
  \end{minipage}%
  \hspace{5mm}
    \begin{minipage}[h]{0.48\textwidth}
    \centering
%
%
\definecolor{mycolor1}{rgb}{0,0,1}%
\definecolor{mycolor2}{rgb}{1,0,0}%
\begin{tikzpicture}

\begin{axis}[%
width=2.3in,
height=2.1in,
at={(0.758in,0.559in)},
scale only axis,
line width=1 pt,
tick label style={font=\boldmath}, 
xmin=-1.8,
xmax=1.8,
xlabel style={font=\color{white!15!black}},
xlabel={\textbf{Normalized I}},
ymin=-1.8,
ymax=1.8,
ylabel style={font=\bfseries\color{white!15!black}},
ylabel={\textbf{Normalized Q}},
axis background/.style={fill=white},
legend style={at={(0.02,0.98)},anchor=north west,legend cell align=left, align=left, draw=white!15!black}
]
\addplot[only marks, mark=*, mark options={}, mark size=1.5000pt, color=mycolor1, fill=mycolor1] table[row sep=crcr]{%
x   y\\
-0.403173082300181  -0.857784521553077\\
0.40648018945488    0.861974264792591\\
0.180845979169001   -0.826869208190699\\
-0.189263535138337  0.834904020511578\\
0.184078768733888   -0.82441750925748\\
-0.188603070037596  0.833077483510358\\
0.183707469054538   -0.823247036750175\\
-0.188064492321096  0.83237719104852\\
0.202957701180663   -0.161903815092728\\
-0.205786712520158  0.162456802923731\\
0.948185304544748   -0.877715545530069\\
-0.947460200128194  0.882693077318887\\
-0.393563192408045  -0.845629248222009\\
0.406415371183562   0.854710752768091\\
-0.991038027502812  -0.940822999663097\\
0.994056733321566   0.940062699146472\\
-0.237614031123236  -0.226958699629777\\
0.236621656049519   0.227931829290281\\
0.393409413158996   0.848823459571183\\
-0.407131993360508  -0.855119038220655\\
-0.196281385122844  0.161234737521016\\
0.202749702190536   -0.165457831266243\\
0.393364570922623   0.849678864755401\\
-0.407503370108663  -0.855594122750581\\
-0.903999729372355  0.195624776276432\\
0.917503422990091   -0.200153250405693\\
-0.993752462594594  -0.940752875240313\\
0.994067305614117   0.94000834270365\\
0.89238355004919    0.369760418986416\\
-0.908884919335297  -0.371126127333964\\
-0.392103827570943  -0.846058578631378\\
0.406344462766873   0.854897081517554\\
0.193964152014954   -0.161547339023181\\
-0.205875644832505  0.162483884570523\\
0.997965453327656   0.940830505110994\\
-0.994081021238369  -0.942274140089252\\
0.905538572524128   -0.193749477695361\\
-0.916442278783314  0.202761715807937\\
-0.392041504301451  -0.845924614651192\\
0.406327672226596   0.854797253163714\\
0.896133052574329   0.369777502162511\\
-0.90955117917372   -0.371257023755\\
0.907472001890895   -0.193936644201101\\
-0.916494504958348  0.202821778833324\\
0.236936860363075   0.227423587974182\\
-0.237945557948998  -0.229854396024031\\
0.201672952393015   -0.162999630891235\\
-0.206704438068926  0.163322185542531\\
0.411630906760254   0.849867951583543\\
-0.407868307309515  -0.856002011074795\\
};
\addlegendentry{\textbf{Raw}}

\addplot[only marks, mark=*, mark options={}, mark size=1.5000pt, color=mycolor2, fill=mycolor2] table[row sep=crcr]{%
x   y\\
0.430583426929023   0.902723170230253\\
-0.434771289860177  -0.905239872694824\\
-0.970617617700797  0.208799790270328\\
0.983440614834358   -0.21273173489318\\
0.963372748600555   -0.205149223301302\\
-0.981033244334231  0.214774211505086\\
-0.235593950801241  0.195032945462722\\
0.249896135658509   -0.199698755313161\\
-0.249310384335577  0.196662434406121\\
0.250586924520102   -0.200515184785609\\
-0.966974432528984  -0.389644255031508\\
0.976157405414604   0.386536188988961\\
-0.289619342276728  -0.272679084757635\\
0.289381468061785   0.272922836884774\\
-0.966058641571048  -0.389438138462399\\
0.975966032736262   0.386460556101076\\
0.419650283861581   0.892257676463652\\
-0.434438880026986  -0.89886200518767\\
-0.24113940309907   0.195396820604569\\
0.249702782179496   -0.199748163021007\\
0.963484889923268   -0.205772778846216\\
-0.981854409689848  0.215132141156409\\
1.00112109495292    0.939513191187956\\
-1.00615653184504   -0.940330474403647\\
0.4356269379991 0.889471202390872\\
-0.434668913802237  -0.897687109152955\\
0.960034571435861   0.387654914863724\\
-0.972052556219356  -0.389525539958268\\
0.962395600104057   0.387780299032382\\
-0.972381237436813  -0.389574976750805\\
-0.961397923177861  0.206025715823681\\
0.981018320685593   -0.211298615511128\\
0.234177501667756   -0.194666755396095\\
-0.252368975899426  0.197417464761684\\
-0.991956243923691  -0.941390276011791\\
1.00652169217945    0.940307183124185\\
0.962579641488031   -0.204328994976842\\
-0.98026899471851   0.214247170339996\\
0.94912288602285    -0.862386223866037\\
-0.95574932126644   0.869095262680828\\
-0.184944412569103  0.862129688168473\\
0.202042917543183   -0.875672813790591\\
-0.433353823961514  -0.888566878320927\\
0.433731256822371   0.898779454629742\\
0.996427500966714   0.93964145289162\\
-1.00596340430301   -0.940238353222344\\
-0.420146366254035  -0.888890174389124\\
0.433665167206499   0.898523580565252\\
0.996283696849972   0.939528264821802\\
-1.00627566105261   -0.940118741521709\\
};
\addlegendentry{\textbf{Pre-distorted}}
 
\end{axis}
 \node[below right, align=left]
at (rel axis cs:1.1,1.13) {\textbf{(1)}};
 \node[below right, align=left]
at (rel axis cs:1.05,0.98) {\textbf{(2)}};
 \node[below right, align=left]
at (rel axis cs:0.87,0.95) {\textbf{(3)}};
 \node[below right, align=left]
at (rel axis cs:0.73,0.93) {\textbf{(4)}};
\end{tikzpicture}%
    \caption{Simulated output 16-QAM constellation of an STHS transmitter array.}
    \label{fig:modifiedQAM}
  \end{minipage}%
\end{figure*}%

\begin{figure*}[h]
    \begin{minipage}[h]{0.48\textwidth}
%
%
\definecolor{mycolor1}{rgb}{1,0,0}%
\definecolor{mycolor2}{rgb}{0,0,1}%

\begin{tikzpicture}
\pgfplotsset{
  grid style = {
    dash pattern = on 0.025mm off 0.95mm on 0.025mm off 0mm, 
    line cap = round,
    gray,
    line width = 1pt
  }
}
\pgfplotsset{set layers}
\begin{axis}[
width=2.4in,
height=2.1in,
scale only axis,
grid=major,
line width=1 pt,
tick label style={font=\boldmath}, 
xmin=20,
xmax=100,
axis y line*=right,
xlabel style={font=\bfseries\color{white!15!black}},
xlabel={\textbf{Frequency (GHz)}},
ymin=30,
ymax=52,
y axis line style={mycolor1,line width=1 pt},
ylabel style={font=\bfseries\color{mycolor1}},
ylabel={\textbf{Output Power (dBm)}},
]
\addplot [color=mycolor1, line width=1.5pt, mark=square, mark options={solid, mycolor1}, forget plot]
  table[row sep=crcr]{%
20	40.5435749056067\\
22	41.2735749056068\\
24	39.6865749056067\\
26	40.6535749056067\\
28	40.8955749056068\\
30	40.5225749056067\\
32	41.3785749056067\\
34	42.2445749056067\\
36	41.7965749056067\\
38	43.7375749056067\\
40	45.9685749056068\\
42	47.4505749056068\\
44	50.4275749056067\\
46	50.4055749056067\\
48	51.2235749056067\\
50	50.2855749056068\\
52	49.6985749056067\\
54	49.3855749056067\\
56	49.2395749056067\\
58	49.3715749056067\\
60	49.5515749056067\\
62	49.7875749056067\\
64	50.1495749056067\\
66	50.3845749056067\\
68	50.6055749056068\\
70	50.8175749056067\\
72	50.8565749056067\\
74	50.8385749056067\\
76	50.6345749056067\\
78	50.2225749056067\\
80	49.6805749056067\\
82	48.9985749056067\\
84	48.2325749056067\\
86	47.5205749056068\\
88	46.8155749056067\\
90	46.0645749056067\\
92	45.4505749056068\\
94	44.8855749056067\\
96	44.2145749056067\\
98	43.7195749056067\\
100	43.3365749056067\\
};
\end{axis}

\begin{axis}[
width=2.4in,
height=2.1in,
scale only axis,
grid=major,
line width=1 pt,
tick label style={font=\boldmath}, 
axis y line*=left,
xmin=20,
xmax=100,
ymin=0,
ymax=30,
y axis line style={mycolor2,line width=1 pt},
ylabel style={font=\bfseries\color{mycolor2}},
ylabel={\textbf{Drain Efficiency (\%)}},
]
\addplot [color=mycolor2, line width=1.5pt, mark=diamond, mark options={solid, mycolor2}, forget plot]
  table[row sep=crcr]{%
20	2.3\\
22	2.7\\
24	1.9\\
26	2.3\\
28	2.4\\
30	2.3\\
32	2.8\\
34	3.3\\
36	3\\
38	4.7\\
40	7.8\\
42	10.9\\
44	21.5\\
46	21.1\\
48	24.9\\
50	20.4\\
52	18.2\\
54	17.3\\
56	16.6\\
58	17.2\\
60	18\\
62	18.9\\
64	20.6\\
66	21.9\\
68	22.9\\
70	24.3\\
72	24.8\\
74	24.8\\
76	24\\
78	21.8\\
80	19.2\\
82	16.4\\
84	13.7\\
86	11.6\\
88	9.8\\
90	8.2\\
92	7.1\\
94	6.2\\
96	5.3\\
98	4.8\\
100	4.3\\
};
\end{axis}
\end{tikzpicture}%
    \caption{Simulated large-signal frequency response of an STHS transmitter array.}
    \label{fig:BW}
  \end{minipage}%
    \hspace{5mm}
  \begin{minipage}[h]{0.48\textwidth}
%
%
\definecolor{mycolor1}{rgb}{0,0,1}%
\definecolor{mycolor4}{rgb}{1,0,0}%
\definecolor{mycolor2}{RGB}{169,169,169}%
\definecolor{mycolor3}{RGB}{128,128,128}%
\begin{tikzpicture}
\pgfplotsset{
  grid style = {
    dash pattern = on 0.025mm off 0.95mm on 0.025mm off 0mm, 
    line cap = round,
    gray,
    line width = 1pt
  }
}
\begin{axis}[%
width=2.6in,
height=2.3in,
at={(0.28\linewidth,0.642in)},
scale only axis,
line width=1 pt,
tick label style={font=\boldmath}, 
xmin=-16,
xmax=0,
xlabel style={font=\bfseries\color{white!15!black}},
xlabel={\textbf{Power Backoff (dB)}},
ymin=0,
ymax=32,
ylabel style={font=\bfseries\color{white!15!black}},
ylabel={\textbf{Drain Efficiency (\%)}},
axis background/.style={fill=white},
xmajorgrids,
ymajorgrids,
legend style={at={(0.02,0.98)},anchor=north west,legend cell align=left, align=left, draw=white!15!black}
]
\addplot [color=mycolor1, line width=1.5pt, mark=diamond, mark options={solid, mycolor1}]
  table[row sep=crcr]{%
-15.8083045978518	3.79848\\
-13.5125383990575	5.4405\\
-11.3609922590417	7.4401\\
-9.21749823582924	10.1234\\
-7.15795976727914	13.36455\\
-5.21444766870727	17.082\\
-3.23272788662766	21.91536\\
-2.08985795555479	23.10399\\
-1.08404765537223	23.51176\\
0	24.24468\\
};
\addlegendentry{\textbf{4-path STHS}}

\addplot [color=mycolor4, line width=1.5pt, mark=square, mark options={solid, mycolor4}]
  table[row sep=crcr]{%
-14.5999719392555   6.2016\\
-12.8208471051384   7.9968\\
-10.3993949055057   11.79\\
-8.35095974833472   15.745\\
-6.67048231494891   19.257\\
-5.51822255922026   20.6295\\
-4.37743202909556   21.918\\
-3.02552400561301   24.3237\\
-2.08762767872174   24.4608\\
-1.00349683467851   25.3455\\
0   25.6553\\
};
\addlegendentry{\textbf{8-path STHS}}

\addplot [color=mycolor2, line width=1.5pt]
  table[row sep=crcr]{%
-15.815	1.62439024390244\\
-15.617	1.71463414634146\\
-15.419	1.80487804878049\\
-15.221	1.89512195121951\\
-15.023	1.98536585365854\\
-14.825	2.07560975609756\\
-14.628	2.16585365853659\\
-14.43	2.25609756097561\\
-14.232	2.34634146341463\\
-14.035	2.43658536585366\\
-13.838	2.61707317073171\\
-13.641	2.70731707317073\\
-13.444	2.79756097560976\\
-13.247	2.9780487804878\\
-13.05	3.06829268292683\\
-12.854	3.24878048780488\\
-12.657	3.3390243902439\\
-12.461	3.51951219512195\\
-12.265	3.7\\
-12.069	3.88048780487805\\
-11.873	4.0609756097561\\
-11.678	4.24146341463415\\
-11.483	4.4219512195122\\
-11.288	4.60243902439024\\
-11.093	4.87317073170732\\
-10.899	5.05365853658537\\
-10.704	5.32439024390244\\
-10.51	5.50487804878049\\
-10.317	5.77560975609756\\
-10.123	6.04634146341463\\
-9.93	6.31707317073171\\
-9.738	6.58780487804878\\
-9.545	6.85853658536585\\
-9.353	7.21951219512195\\
-9.162	7.49024390243902\\
-8.971	7.85121951219512\\
-8.78	8.21219512195122\\
-8.59	8.57317073170732\\
-8.4	8.93414634146341\\
-8.211	9.29512195121951\\
-8.022	9.74634146341463\\
-7.834	10.1073170731707\\
-7.647	10.5585365853659\\
-7.46	11.009756097561\\
-7.275	11.4609756097561\\
-7.09	11.9121951219512\\
-6.906	12.3634146341463\\
-6.723	12.9048780487805\\
-6.541	13.3560975609756\\
-6.361	13.8975609756098\\
-6.182	14.4390243902439\\
-6.005	14.890243902439\\
-5.829	15.4317073170732\\
-5.656	15.9731707317073\\
-5.485	16.5146341463415\\
-5.315	17.0560975609756\\
-5.149	17.5975609756098\\
-4.985	18.0487804878049\\
-4.823	18.590243902439\\
-4.664	19.0414634146341\\
-4.507	19.4926829268293\\
-4.353	20.0341463414634\\
-4.201	20.3951219512195\\
-4.051	20.8463414634146\\
-3.903	21.2975609756098\\
-3.756	21.6585365853659\\
-3.61	22.019512195122\\
-3.465	22.380487804878\\
-3.321	22.7414634146341\\
-3.178	23.1024390243902\\
-3.035	23.3731707317073\\
-2.893	23.7341463414634\\
-2.751	24.0048780487805\\
-2.61	24.3658536585366\\
-2.469	24.6365853658537\\
-2.328	24.9073170731707\\
-2.188	25.1780487804878\\
-2.048	25.4487804878049\\
-1.909	25.7195121951219\\
-1.77	25.990243902439\\
-1.631	26.2609756097561\\
-1.492	26.5317073170732\\
-1.354	26.8024390243902\\
-1.217	27.0731707317073\\
-1.079	27.3439024390244\\
-0.942	27.6146341463415\\
-0.805	27.8853658536585\\
-0.668999999999999	28.1560975609756\\
-0.533999999999999	28.4268292682927\\
-0.398999999999999	28.6975609756098\\
-0.264999999999999	29.0585365853659\\
-0.132	29.3292682926829\\
0	29.6\\
};
\addlegendentry{\textbf{Doherty PA}}
\addplot [color=mycolor3,dashed, line width=1.5pt]
  table[row sep=crcr]{%
-15.8083045978518	0.777071983598035\\
-13.5125383990575	1.3183716976032\\
-11.3609922590417	2.16367728211257\\
-9.21749823582924	3.54439314119237\\
-7.15795976727914	5.69502630408154\\
-5.21444766870727	8.90936895227219\\
-3.23272788662766	14.0610878755604\\
-2.08985795555479	18.2938837713116\\
-1.08404765537223	23.0614678008053\\
0	29.6\\
};
\addlegendentry{\textbf{Class A/B}}

\end{axis}

\end{tikzpicture}%
    \caption{Simulated continuous wave large-signal characteristics  of an STHS transmitter array.}
    \label{fig:Eff_comparison}
\end{minipage}%
\end{figure*}%



\subsection{STHS Array Beamforming Result}
The spatial beamforming result with the first harmonic targeted at $\theta = 20^{\circ}$ is simulated.
The circuit-generated radiation patterns for STHS array with different duty cycle ratio $\alpha$ are shown in Figs. \ref{fig:simulated0}, \ref{fig:simulated3}. It is seen that at 6-dB PBO, the $m = -3$ harmonic is suppressed similar to the ideal Matlab simulation. 
The simulated harmonic efficiency $\zeta_{harm}$
 at different power back-off regions are compared with the ideal values in Fig. \ref{fig:TMAeff}. It is seen that the harmonic efficiency is kept above 80\% beyond the 3-dB PBO due to the elimination of $m = -3$ harmonic. In the deep power back-off region, the relative power of other undesired harmonics will increase, which degrades the harmonic efficiency. The simulated harmonic efficiency shows an approximately 5\% lower harmonic efficiency compared to the ideal value which can be attributed to the nonidealities of the time-modulated PA such as the switching overshoot, leakage power, etc. 

\subsection{STHS Array Back-off Result}
Fig. \ref{fig:aVsPower} shows how the output power of the desired first harmonic can be controlled by the duty cycle ratio of the modulation pulse $\alpha$.
Since the power of the first harmonic also depends on the harmonic efficiency, the slope of the output power is steeper for $10log(\alpha)<-3$ region where the harmonic efficiency begins to degrade quicker.

 The illustration of amplitude and phase for a 16 QAM is shown in Fig. \ref{fig:qamsignal}. The STHS array works in the peakmode with a duty cycle ratio $10log(\alpha)=0$ to generate  symbol 1 $(3+3j)$.
As for symbol 3 ($1+1j$), the IF signal will remain the same as the signal for symbol 1 while the duty cycle ratio for the modulating pulses is adjusted to be $10log(\alpha) = -4.8$. To generate symbol 4 ($-1+1j$), the I channel changes its polarity from positive to negative while the Q channel and the duty cycle ratio remain the same as the signal for symbol 3. Furthermore, since the power comes in the RF power amplifier should remain the maximum power for all signals transmitted to achieve high efficiency, to generate the symbol 2 ($3+1j$), the signal for I channel and Q channel is designed such that a combination of I/Q channel will result in the maximum power for power amplifier and the angle on the constellation should be $arctan(\frac{1}{3})$ as required.
The simulated output constellation of the STHS array for raw signals and pre-distorted signals are both shown in Fig. \ref{fig:modifiedQAM}.
Since the output power on the first harmonic is affected by both harmonic efficiency and circuit efficiency of the STHS array, the amplitude of the generated signal will be distorted if the efficiency degeneration of circuit and array is not taken into consideration when choosing the duty cycle ratio to deliver back-off signal. To have better control of the amplitude of output signal power, the duty cycle ratio should be bigger than the ideal case if the back-off signal is delivered. After modification, an improvement in output power control is observed.

The maximum PA output power and the peak drain efficiency are simulated by sweeping the frequency of the input signal (Fig. \ref{fig:BW}). The continuous wave (CW) large-signal characteristics of the STHS transmitter array are simulated and plotted in Fig. \ref{fig:Eff_comparison}. 
The peak STHS array DE is 24.2\% while a typical class-A/B PA has a drain efficiency of 29.6\%. This is contributed by both the efficiency loss of the switching circuit and the existence of higher harmonics.  The drain efficiency of the STHS transmitter array is 22\% at 3-dB power back-off. This back-off efficiency is 57\% larger than a class-A/B amplifier with the same peak drain efficiency. The array has a DE of 16\% at 6-dB PBO, showing a 100\% improvement compared to a class-A/B. The DE of the STHS array is as high as 10.2\% at 9-dB power back-off. This back-off efficiency exhibits a 190\% improvement compared to a class-A/B amplifier. The STHS array is also compared with a Doherty PA with normalized drain efficiency. It is seen that the STHS behaves better than the Doherty PA in the deep PBO region of >5-dB. Compared with a Doherty PA, the STHS array DE shows a 12\% efficiency enhancement at 7-dB PBO, and a 33\% enhancement at 9-dB PBO. Moreover, at the deep PBO region of 13-dB PBO, the array has a DE of 5.9\%, showing a 96\% improvement compared with a Doherty PA. 

It is worth mentioning that if the size of the footprint is not the major design constraint, the 8-path STHS array structure can be utilized to realize a further back-off efficiency enhancement. At 6-dB PBO, an 8-path STHS array achieves a DE of 20\% which shows a 35\% improvement compared with a Doherty PA and a 170\% improvement compared with a class-A/B. At 10-dB PBO, an 8-path STHS array achieves a DE of 12\%. This DE has a 99\% improvement compared with a Doherty PA and a 290\% enhancement compared with a class-A/B. 

\section{Conclusion}
In this paper, we presented an EM-circuit co-designed and jointly optimized mm-wave transmitter array for back-off efficiency enhancement. 

We proposed a 4-path STHS array where the duty cycle of the modulating pulses is utilized to control power generation
while other pulse parameters are designed to suppress undesired harmonics in the power back-off region. The 8-path STHS array is also designed while it is favored if the size of the footprint is not the major design constraint to have a further back-off efficiency enhancement.

The stacking time-modulated PA structure is proposed to realize the pulse modulation, and   
it's accurate circuit power and efficiency model with an equivalent switching capacitor is proposed to analyze the energy flow resulting from dynamic switching loss and static leakage current. 
Compared with the existing structure involving SPST switches along with the signal path, this time-modulated PA structure reduces the insertion loss when the switch is "ON" and allows smaller power leakage when the switch is "OFF".

The design and analysis is validated through the simulation of a two-stage power amplifier simulation in 65-nm CMOS at 77 GHz.
The 4-path STHS array DE is 57\% larger at 3-dB PBO, 100\% larger at 6-dB PBO, and 190\% larger at 9-dB PBO compared with class-A/B PA. 
As for the 8-path STHS array, it achieves a 170\% DE improvement at 6-dB PBO, and a 290\% DE improvement at 10-dB PBO compared with class-A/B PA. 

The STHS transmitter array originally adds spatial and temporal dimensions to realize back-off efficiency enhancement.
Furthermore, it is worth emphasizing the compatibility of this method with other existing back-off enhancement structures such as the Doherty amplifiers and  envelope tracking to achieve additional enhancement.  
This will be one potential solution to deal with  device-limited back-off efficiency.

\bibliographystyle{IEEEtran}
\bibliography{references}  

\begin{thebibliography}{10}
\providecommand{\url}[1]{#1}
\csname url@samestyle\endcsname
\providecommand{\newblock}{\relax}
\providecommand{\bibinfo}[2]{#2}
\providecommand{\BIBentrySTDinterwordspacing}{\spaceskip=0pt\relax}
\providecommand{\BIBentryALTinterwordstretchfactor}{4}
\providecommand{\BIBentryALTinterwordspacing}{\spaceskip=\fontdimen2\font plus
\BIBentryALTinterwordstretchfactor\fontdimen3\font minus
  \fontdimen4\font\relax}
\providecommand{\BIBforeignlanguage}[2]{{%
\expandafter\ifx\csname l@#1\endcsname\relax
\typeout{** WARNING: IEEEtran.bst: No hyphenation pattern has been}%
\typeout{** loaded for the language `#1'. Using the pattern for}%
\typeout{** the default language instead.}%
\else
\language=\csname l@#1\endcsname
\fi
#2}}
\providecommand{\BIBdecl}{\relax}
\BIBdecl

\bibitem{afaqui2016ieee}
M.~S. Afaqui, E.~Garcia-Villegas, and E.~Lopez-Aguilera, ``{IEEE 802.11 ax}:
  Challenges and requirements for future high efficiency {WiFi},'' \emph{IEEE
  Wireless Communications}, vol.~24, no.~3, pp. 130--137, 2016.

\bibitem{cao2020pseudo}
Y.~Cao and K.~Chen, ``{Pseudo-Doherty} load-modulated balanced amplifier with
  wide bandwidth and extended power back-off range,'' \emph{IEEE Transactions
  on Microwave Theory and Techniques}, vol.~68, no.~7, pp. 3172--3183, 2020.

\bibitem{kazimierczuk2008rf}
M.~K. Kazimierczuk, \emph{{RF} power amplifiers}.\hskip 1em plus 0.5em minus
  0.4em\relax Wiley Online Library, 2008, vol.~1.

\bibitem{cripps2006rf}
S.~C. Cripps, \emph{{RF} power amplifiers for wireless communications}.\hskip
  1em plus 0.5em minus 0.4em\relax Artech house Norwood, MA, 2006, vol.~2.

\bibitem{chowdhury2009design}
D.~Chowdhury, P.~Reynaert, and A.~M. Niknejad, ``Design considerations for 60
  {GHz} transformer-coupled {CMOS} power amplifiers,'' \emph{IEEE Journal of
  Solid-State Circuits}, vol.~44, no.~10, pp. 2733--2744, 2009.

\bibitem{shanks1959four}
H.~Shanks and R.~Bickmore, ``Four-dimensional electromagnetic radiators,''
  \emph{Canadian Journal of Physics}, vol.~37, no.~3, pp. 263--275, 1959.

\bibitem{poli2011harmonic}
L.~Poli, P.~Rocca, G.~Oliveri, and A.~Massa, ``Harmonic beamforming in
  time-modulated linear arrays,'' \emph{IEEE Transactions on Antennas and
  Propagation}, vol.~59, no.~7, pp. 2538--2545, 2011.

\bibitem{bregains2008signal}
J.~Bregains, J.~Fondevila-Gomez, G.~Franceschetti, and F.~Ares, ``Signal
  radiation and power losses of time-modulated arrays,'' \emph{IEEE
  Transactions on Antennas and Propagation}, vol.~56, no.~6, pp. 1799--1804,
  2008.

\bibitem{yang2005design}
S.~Yang, Y.~B. Gan, A.~Qing, and P.~K. Tan, ``Design of a uniform amplitude
  time modulated linear array with optimized time sequences,'' \emph{IEEE
  Transactions on Antennas and Propagation}, vol.~53, no.~7, pp. 2337--2339,
  2005.

\bibitem{fondevila2004optimizing}
J.~Fondevila, J.~Bregains, F.~Ares, and E.~Moreno, ``Optimizing uniformly
  excited linear arrays through time modulation,'' \emph{IEEE Antennas and
  Wireless Propagation Letters}, vol.~3, no.~1, pp. 298--301, 2004.

\bibitem{poli2010handling}
L.~Poli, P.~Rocca, L.~Manica, and A.~Massa, ``Handling sideband radiations in
  time-modulated arrays through particle swarm optimization,'' \emph{IEEE
  Transactions on Antennas and Propagation}, vol.~58, no.~4, pp. 1408--1411,
  2010.

\bibitem{zhu2012design}
Q.~Zhu, S.~Yang, L.~Zheng, and Z.~Nie, ``Design of a low sidelobe time
  modulated linear array with uniform amplitude and sub-sectional optimized
  time steps,'' \emph{IEEE Transactions on Antennas and Propagation}, vol.~60,
  no.~9, pp. 4436--4439, 2012.

\bibitem{li2010hybrid}
G.~Li, S.~Yang, Y.~Chen, and Z.~Nie, ``A hybrid analog-digital adaptive
  beamforming in time-modulated linear arrays,'' \emph{Electromagnetics},
  vol.~30, no.~4, pp. 356--364, 2010.

\bibitem{li2010direction}
G.~Li, S.~Yang, and Z.~Nie, ``Direction of arrival estimation in time modulated
  linear arrays with unidirectional phase center motion,'' \emph{IEEE
  Transactions on Antennas and Propagation}, vol.~58, no.~4, pp. 1105--1111,
  2010.

\bibitem{masotti2016time}
D.~Masotti, A.~Costanzo, M.~Del~Prete, and V.~Rizzoli, ``Time-modulation of
  linear arrays for real-time reconfigurable wireless power transmission,''
  \emph{IEEE Transactions on Microwave Theory and Techniques}, vol.~64, no.~2,
  pp. 331--342, 2016.

\bibitem{yao2015single}
A.-M. Yao, W.~Wu, and D.-G. Fang, ``Single-sideband time-modulated phased
  array,'' \emph{IEEE Transactions on Antennas and Propagation}, vol.~63,
  no.~5, pp. 1957--1968, 2015.

\bibitem{kang2013envelope}
D.~Kang, B.~Park, D.~Kim, J.~Kim, Y.~Cho, and B.~Kim, ``{Envelope-tracking CMOS
  power amplifier module for LTE applications},'' \emph{IEEE Transactions on
  Microwave Theory and Techniques}, vol.~61, no.~10, pp. 3763--3773, 2013.

\bibitem{mahmoudidaryan2019wideband}
P.~Mahmoudidaryan, D.~Mandal, B.~Bakkaloglu, and S.~Kiaei, ``Wideband hybrid
  envelope tracking modulator with hysteretic-controlled three-level switching
  converter and slew-rate enhanced linear amplifier,'' \emph{IEEE Journal of
  Solid-State Circuits}, vol.~54, no.~12, pp. 3336--3347, 2019.

\bibitem{kim2006doherty}
B.~Kim, J.~Kim, I.~Kim, and J.~Cha, ``The {Doherty} power amplifier,''
  \emph{IEEE microwave magazine}, vol.~7, no.~5, pp. 42--50, 2006.

\bibitem{camarchia2015doherty}
V.~Camarchia, M.~Pirola, R.~Quaglia, S.~Jee, Y.~Cho, and B.~Kim, ``The
  {Doherty} power amplifier: Review of recent solutions and trends,''
  \emph{IEEE Transactions on Microwave Theory and Techniques}, vol.~63, no.~2,
  pp. 559--571, 2015.

\bibitem{kaymaksut2015transformer}
E.~Kaymaksut, D.~Zhao, and P.~Reynaert, ``Transformer-based {Doherty power
  amplifiers} for mm-wave applications in 40-nm {CMOS},'' \emph{IEEE
  Transactions on Microwave Theory and Techniques}, vol.~63, no.~4, pp.
  1186--1192, 2015.

\bibitem{raab1985efficiency}
F.~Raab, ``Efficiency of outphasing {RF} power-amplifier systems,'' \emph{IEEE
  Transactions on Communications}, vol.~33, no.~10, pp. 1094--1099, 1985.

\bibitem{godoy20122}
P.~A. Godoy, S.~Chung, T.~W. Barton, D.~J. Perreault, and J.~L. Dawson, ``A
  {2.4-GHz, 27-dBm asymmetric multilevel outphasing power amplifier in 65-nm
  CMOS},'' \emph{IEEE Journal of Solid-State Circuits}, vol.~47, no.~10, pp.
  2372--2384, 2012.

\bibitem{tai2012transformer}
W.~Tai, H.~Xu, A.~Ravi, H.~Lakdawala, O.~Bochobza-Degani, L.~R. Carley, and
  Y.~Palaskas, ``A transformer-combined 31.5 {dBm outphasing power amplifier in
  45 nm LP CMOS} with dynamic power control for back-off power efficiency
  enhancement,'' \emph{IEEE Journal of Solid-State Circuits}, vol.~47, no.~7,
  pp. 1646--1658, 2012.

\bibitem{zhang2019subharmonic}
A.~Zhang and M.~S.-W. Chen, ``A subharmonic switching digital power amplifier
  for power back-off efficiency enhancement,'' \emph{IEEE Journal of
  Solid-State Circuits}, vol.~54, no.~4, pp. 1017--1028, 2019.

\end{thebibliography}

\end{document}